\documentclass[apj,twocolumn,twocolappendix,numberedappendix]{emulateapj}
\usepackage{amsmath}
\usepackage{CJKutf8}
\newcommand{\mum}{\ifmmode{\rm \mu m}\else{$\mu$m}\fi}

\newcommand{\spitzer}{{\em Spitzer}\ }
\newcommand{\spitzerirs}{{\em Spitzer}/IRS\ }
\newcommand{\herschel}{{\em Herschel}\ }

\newcommand{\metalOH}{12+log(O/H)}

\newcommand{\cosmology}{$\Omega _m = 0.27$,~$\Omega_{\Lambda}=0.73$~and~$H_0=71$km~s$^{-1}$Mpc$^{-1}$}

\newcommand{\zgtrsim}[1]{\ensuremath{z\gtrsim{#1}}}

\slugcomment{Accepted by ApJ \today}

\begin{document}

\title{The contribution of host galaxies to the infrared energy output of \zgtrsim{5.0} quasars}
\author{
    Jianwei Lyu (\begin{CJK}{UTF8}{gbsn}吕建伟\end{CJK})\altaffilmark{1,\dag},
    G. H. Rieke\altaffilmark{1},
    Stacey Alberts\altaffilmark{1}
}

\altaffiltext{1}{Steward Observatory, University of Arizona,
   933 North Cherry Avenue, Tucson, AZ 85721, USA}

\altaffiltext{\dag}{jianwei@email.arizona.edu}

\begin{abstract}
    The infrared spectral energy distributions (SEDs) of $z\gtrsim 5$ quasars
    can be reproduced by combining a low-metallicity galaxy template with a
    standard AGN template.  The host galaxy is represented by Haro 11, a
    compact, moderately low metallicity, star-bursting galaxy that shares
    typical features of high-$z$ galaxies. For the vast majority of $z\gtrsim
    5$ quasars, the AGN contribution is well modeled by a standard empirical
    template with the contamination of star formation in the infrared
    subtracted.  Together, these two templates can separate the contributions
    from the host galaxy and the AGN even in the case of limited data points,
    given that this model has only two free parameters. Using this method, we
    re-analyze 69 $z\gtrsim 5$ quasars with extensive \herschel observations,
    and derive their AGN luminosities $L_{\rm AGN}$ in a range $\sim
    (0.78-27.4) \times10^{13}\, L_{\odot}$, the infrared luminosities from star
    formation $L_{\rm SF,IR} \sim (<1.5-25.7)\times10^{12}\, L_{\odot}$, and
    the corresponding star formation rates ${\rm SFR}\sim (<290-2650)\,
    M_\odot/{\rm yr}$. The average infrared luminosity from star formation and
    the average total AGN luminosity of the $z\gtrsim5$ quasar sample follows
    the correlation defined by quasars at $z < 2.6$.  We assume these quasar
    host galaxies maintain a constant average SFR ($\sim620\, M_\odot/{\rm
    yr}$) during their mass assembly and estimate the stellar mass that could
    form till $z\sim5-6$ to be $\langle M_* \rangle \sim(3-5)\times10^{11}
    M_\odot$. Combining with the black hole (BH) mass measurements, this
    stellar mass is adequate to establish a BH-galaxy mass ratio $M_{\rm
    BH}/M_{*}$ at 0.1-1\%, consistent with the local relation.
\end{abstract}

\keywords
{galaixies: active -- infrared: galaxies -- quasars: general -- galaxies: dwarf }

\section{Introduction}

Over 150 quasars with black hole masses of the order of $10^8-10^9\,M_{\odot}$
have been discovered beyond $z\sim5$, less than a billion years after the Big
Bang \citep[e.g.,][]{Fan2006, Jiang2008, Mortlock2009, Willott2010a,
Morganson2012, McGreer2013, Banados2014}. In the local Universe, the masses of
supermassive black holes (SMBHs) are correlated with the properties of their
host galaxies, suggesting galaxies and SMBHs possibly coevolve \citep[][and
references therein]{Kormendy-Ho2013}.  Since star formation (SF) enables the
buildup of galaxies and active galactic nuclei (AGN) trace the growth of SMBHs,
the so-called SF-AGN relation has come under intense scrutiny for decades
\citep[][and references therein]{Heckman-Best2014}.  With the use of ground-
and space-based facilities, similar research can be extended to $z > 5$,
allowing the preliminary examination of both the stellar and SMBH growth at
very early stages in galaxy evolution \citep[e.g.,][]{Walter2004, Walter2009,
Maiolino2005, Riechers2006, Jahnke2009, Wang2010, Wang2013, Willott2013,
Willott2015}.

However, observing the AGN host galaxy in a quasar is challenging, since the
bright continuum emission from the active nucleus overpowers the light from the
galaxy from the UV through the near-infrared (near-IR or NIR). At $z > 5$, the
situation is even more extreme: the AGN tend to be the most luminous because of
the evolution of the quasar luminosity function and selection effects
\citep[e.g.,][]{Fan2004,Jiang2008,Willott2010b,McGreer2013}; the host galaxies
are found to be compact and small \citep[e.g.,][]{Jiang2013, Wang2013}; and the
presence of copious amounts of dust \citep[e.g.,][]{Wang2008a,Leipski2014} may
also obscure the galaxy light. To study these very distant AGN host galaxies,
attention has turned to the emission in longer wavelengths
\citep[e.g.,][]{Wang2010,Wang2011a,Wang2013,Leipski2013,Leipski2014}, where the
AGN is less dominant.

A promising tool to probe the stellar activity in a quasar is analyzing its
infrared (IR) spectral energy distribution (SED). For galaxies, most of the
radiation from the recently formed stars is absorbed and re-emitted at IR
wavelengths. However, in quasars, the central AGN may also heat the dust
\citep[e.g.,][]{Haas2003, Netzer2007}. Since spatially resolving the IR
emission at high-redshift is impossible with current facilities, we have to
rely on SED models to distinguish star formation and nuclear activity (e.g.,
\citealt{Marshall2007,Mullaney2011,Mor-Netzer2012,Magdis2013, Netzer2014, Xu2015a, Xu2015b}).

The UV-to-NIR SEDs of AGN seem to have little evolution with redshift and
Eddington-ratio (e.g., \citealt{Hao2014, Wang2008a}). However, at $z > 5$, the
far-IR SEDs of quasars seem to include a warm (40-60 K) dust component
\citep[e.g.,][]{Beelen2006, Wang2008b, Leipski2014}, which is not commonly
found in the local quasars. It is intriguing to check if such behavior is
due to the evolution of the AGN host galaxies, since the IR SEDs of galaxies do
change at high-$z$ (see \citealt{Lutz2014} for a review).  Compared with local
nuclear-concentrated ULIRGs, intermediate redshift ($z\sim2-3$) dusty
star-forming galaxies (DSFGs) are more extended, resulting in cooler SEDs than
those locally with similar IR luminosities \citep{Rujopakarn2011}.  Meanwhile,
galaxies at higher redshifts have more gas \citep[e.g.,][]{Carilli-Walter2013}
to supply the star formation, boosting their IR luminosities
\citep[e.g.,][]{Daddi2005, Daddi2007, Rodighiero2011, Magnelli2014,
Schreiber2015}. Moreover, both observations and theories suggest galaxies in
the early Universe are generally metal poor \citep[e.g.,][and references
therein]{Madau-Dickinson2014}, which could also have detectable effects on
their SEDs, such as weak aromatic features, featureless mid-IR (MIR) continuum,
and higher dust temperatures \citep[e.g.,][]{Galliano2005,Remy-Ruyer2013}.  At
extremely high redshift (e.g., $z>4$), direct constraints on metallicity in
galaxies are rare; however, indirect evidence supporting low-metallicity comes
from the failed detection of the dust continuum for nearly all $z>6$ galaxies
(e.g., \citealt{Ouchi2013, Maiolino2015}, but see \citealt{Watson2015}). It is likely that
galaxies, in general, including those quasar host galaxies, are of relatively
low-metallicity when the Universe age is within $\sim$ 1~Gyr.

Because of these issues, fits to the SEDs of high redshift quasars are
unsuccessful using conventional quasar templates plus those for typical normal
(e.g., $\sim$ solar metallicity) SF galaxies, a method that works well for
low-to-intermediate redshift quasars (e.g., \citealt{Mullaney2011,Magdis2013,
Xu2015a}). In this paper, we demonstrate that the SEDs of $z\gtrsim5$ quasars
can be modeled using a moderately low-metallicity galaxy template to represent
the AGN host galaxy. We combine a galaxy template derived from Haro 11 and a
modified AGN template based on \cite{Elvis1994} to provide physically-motivated
fits that successfully reproduce these $z\gtrsim 5$ quasar infrared SEDs. This
simple model can be used to probe the relation between the AGN activity and
host star formation in quasars with very limited observations.

Throughout this paper, we adopt cosmology \cosmology.

\section{Selection of SED templates}\label{sec:templates}

 In modeling the SEDs of quasars at $z\gtrsim$ 5.0, the dearth of data points
 at long wavelengths requires minimizing the number of free
 parameters in SED fitting.  While more precise fittings may be achieved by
 adding more free parameters, the scientific interpretation is then more
 subject to model degeneracy. To first order, the SED of a quasar should
 consist of an AGN component, and a host galaxy component.  If suitable SED
 templates can be found, we only need two free parameters to normalize their
 contributions. Such a two-parameter model can be used to retrieve information
 from sources with less complete observations and make the interpretation less
 ambiguous.

\subsection{Host Galaxy Template: Why Haro 11?}\label{sec:host-template}

Galaxies at $z\sim2-3$ are of relatively low-metallicity
\citep[e.g.,][]{Cullen2014, Maier2014}. Confirming the trend toward lower
metallicity to $z\gtrsim5$ is difficult with current capabilities. Recently,
some groups have tried to detect the dust continua of  $z>6$ normal galaxies
using the Atacama Large Millimeter/Sub-millimeter Array (ALMA).  The unexpected
failures of almost all of these efforts have led to the interpretation that
these $z>6$ galaxies may be scaled-up versions of local very metal-poor dwarf
galaxies \citep[e.g.,][]{Ouchi2013, Maiolino2015}. As
\cite{Fisher2014} pointed out, it would be almost impossible to observe the
dust emission of any $z>6$ galaxies with extreme low metallicity like the local
dwarf galaxy 1 Zw 18. For the most luminous systems, however, the change in
metallicity seems to be modest \citep[e.g.,][]{Rawle2014}.  As a result, the
ideal sample to draw a representative template for IR observable high-$z$
galaxies is the moderately low-metallicity galaxies in the local Universe.

In addition,  AGN host galaxies at $z > 4$ are found to be compact with typical
sizes $\sim$ 1-3 kpc, from observations at rest-frame UV \citep{Jiang2013},
deep $K_s$-band images \citep{Targett2012}, dust continuum maps
\citep{Wang2013}, submm fine structure line maps
\citep[e.g.,][]{Wang2013, Willott2013, Willott2015}, molecular gas distributions
\citep[e.g.,][]{Walter2004, Walter2009, Wang2013}, and from SED analysis
\citep{Greve2012}. Compared with extended galaxies of the same infrared
luminosity, they are expected to have hotter far-IR SEDs due to compact star
forming regions \citep{Groves2008}.  Thus, we are motivated to search for a
moderately low-metallicity galaxy with a high surface density of star formation
to provide a SED analogous to that we expect for the star formation in the host
galaxies of high-$z$ quasars.

Appendix \ref{sec:low-metal-template} presents the procedure to derive
low-metallicity galaxy templates. To summarize briefly, we began with the
sample of the Dwarf Galaxy Survey (DGS; \citealt{Madden2013}), which includes
the largest metallicity range observable in the local Universe, with
\metalOH~ranging from 7.14 to 8.43, and spans four orders of magnitude in star
formation rates. Combining their \herschel far-IR data \citep{Remy-Ruyer2013}
and archival WISE mid-IR photometry, we fit the broad-band SEDs with a far-IR
modified blackbody plus a mid-IR power-law component, and replaced the mid-IR
fit SEDs with the corresponding \spitzer spectra. Among the 19 dwarf galaxies
studied in detail, Haro 11 is the best candidate analog for high-z galaxies.
Haro 11 is a moderately low-metallicity ($Z=1/3\,Z_{\odot}$,
\citealt{James2013}) dwarf ($M_{*}=10^{10}\,M_{\odot}$, \citealt{Ostlin2001})
galaxy in the nearby Universe ($D=92.1$ Mpc, \citealt{Bergvall2006}). It shows
substantial star formation activity (${\rm SFR}\approx 20-30\,M_{\odot}/{\rm yr}$,
\citealt{Grimes2007}, see also Appendix~\ref{sec:dwarf-sf-haro11}) and emits
strongly in the infrared ($L_{\rm IR} \approx 2.0\times10^{11}\, L_\odot$,
\citealt{Adamo2010}). Haro 11 also contains an extremely young stellar
population with age~$<40$ Myr \citep{Adamo2010}. Some authors suggest it is a
local analogue of the high-$z$ Lyman break galaxies (LBGs) or Lyman-$\alpha$
emitters \citep{Hayes2007,Leitet2011}.

\begin{figure}[htp]
    \begin{center} 
	\includegraphics[width=1.0\hsize]{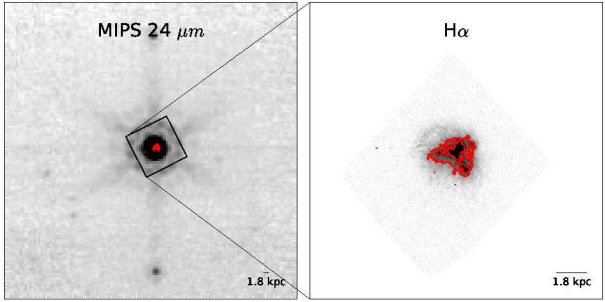} 
	\caption{ 
	    The \spitzer MIPS 24$\mum$ (left; Program ID: 59, PI: G. H.
	    Rieke) and the HST H$\alpha$ (right; \citealt{Ostlin2009}) images
	    of Haro 11. The H$\alpha$ image is zoomed to view the central
	    square region on the 24$\mum$ image.  We also overplot the same
	    H$\alpha$ contours (red lines) on both images.
	}
	\label{fig:haro11_image} 
    \end{center} 
\end{figure}

Besides low metallicity, the most important two features of Haro 11 are its
high star formation rate and compact size, indicating a very high star
formation surface density. From our estimation, the star formation rate of Haro
11 can be as high as $\sim32\,M_\odot/{\rm yr}$ (based on $L_{\rm IR}$ and
$L_{\rm FUV}$, see Appendix~\ref{sec:dwarf-sf-haro11}), which is
significantly higher than the vast majority of dwarf galaxies in the literature
\citep{Hopkins2002}.  Meanwhile, Haro 11 has a compact size. Its MIPS 24$\mum$
image is perfectly diffraction-limited (see Figure~\ref{fig:haro11_image}),
which puts an upper-limit on its IR emitting region size ($< 3.4''$ or 1.2
kpc). The size of the star formation region of Haro 11 constrained from
high-resolution H$\alpha$ images \citep{Ostlin2009} is also small ($\sim1.3$
kpc from measuring 50\% total flux, and $\sim$2.7 kpc from measuring 90\% total
flux). The IR luminosity surface density, $\Sigma _{L({\rm IR})}$, of Haro 11
is $\sim 10^{11}\, L_\odot/{\rm kpc}^2$, which approaches the values in
galaxies at $z\gtrsim4$ (e.g., GN20 has $\Sigma _{L({\rm IR})}\sim
10^{12}\,L_\odot/{\rm kpc}^2$, \citealt{Hodge2015}).  The high star formation
rate surface density and infrared luminosity surface density of Haro 11 are
exceptional among dwarf galaxies, making it the most suitable local analog to
high-$z$ quasar host galaxies.

\begin{figure}[htp]
    \begin{center} 
	\includegraphics[width=0.8\hsize]{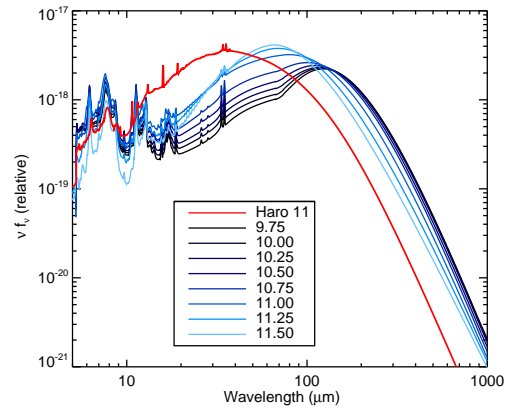} 
	\caption{ 
	    Comparison of Haro 11 template (red line) and the normal
	    star-forming galaxy templates with $\log L_{\rm IR} =
	    9.75\sim11.50$ in \cite{Rieke2009} (blue lines). All the templates
	    are normalized to have the same $L_{\rm IR,template}$
	}
	\label{fig:dwarf_template} 
    \end{center} 
\end{figure}

In Figure~\ref{fig:dwarf_template}, we compare the derived Haro 11 template
with a number of normal solar-metallicity star-forming (SF) templates in
\cite{Rieke2009}. Haro 11 has a larger mid-IR slope with $\alpha=3.63$
($f_\nu\propto\lambda^\alpha$), in contrast with normal galaxies with
$\alpha\sim2.0$ \citep[e.g.,][]{Blain2003,Casey2012}. The derived dust
temperature is T = 46.5 K with emissivity index $\beta=1.9$.  Haro 11 also
presents very weak aromatic features compared with normal galaxies.  All these
characteristics are commonly seen for other dwarf galaxies (see
Appendix~\ref{sec:dwarf-template-properties}). Compared with the \cite{Rieke2009} SED
templates with $\log L_{\rm IR} < 11.50$, which are representative for $z\sim2$
galaxies (see Section~\ref{sec:test-galaxy}), Haro 11 has similar $L_{\rm IR}$
surface densities but higher dust temperature. We suggest the low-metallicity
of Haro 11 is the most likely reason for its warmer SED.

\subsection{AGN Continuum Template}\label{sec:AGN-template}

Candidate AGN templates can be derived from either numerical models
\citep[e.g.,][]{Fritz2006,Honig-Kishimoto2010} or semi-analytic models
\citep[e.g.,][]{Mullaney2011,Sajina2012}. However, such models always have many
free parameters, which need to be optimized to fit real AGN behavior. 
Hence the starting point for determining AGN templates needs to be an accurate
empirical version.  

\cite{Elvis1994} built an X-ray to radio SED template for a sample of 47
well-defined optically selected quasars and subtracted the host galaxy emission
in the UV/optical and near-IR bands. This template has become the classic
representation of Type 1 AGN SEDs in the ultraviolet, visible, near, and
mid-infrared.  Many studies based on larger samples and modern data have
closely reproduced the Elvis template \citep[e.g.,][]{Richards2006, Shang2011,
Runnoe2012a, Elvis2012, Hanish2013, Scott2014}. The remarkable similarity of
these results to the Elvis template is demonstrated by the comparisons in
\cite{Scott2014} (their Figure 5). The success of the Elvis template is also
demonstrated by its broad application, for example, to study the SEDs of type-1
AGN in XMM-COSMOS \citep{Elvis2012} and decompose the SEDs of
intermediate-redshift quasars in \cite{Xu2015a}. The template shape appears to
vary little with cosmic evolution or other characteristics such as the
Eddington ratio \citep[e.g.,][]{Hao2011, Hao2014}.  In particular, this
template appears to work equally well to z $\sim$ 6. \cite{Jiang2006}
demonstrated that the rest-frame $0.15-3.5\mum$ SEDs of 13 z $\sim$ 6 quasars
can be matched with the \cite{Elvis1994} template. \cite{Wang2008a}
demonstrated that the average optical-to-near-IR SED of 33 z $\sim$ 6
quasars is consistent with the Elvis template. \cite{Jiang2010} find that the
near infrared and optical-to-NIR colors of hundreds of quasars are virtually
the same from the local epoch to z $\ge$ 6, i.e., they are consistent with a
common SED shape, which must therefore be consistent with the Elvis template.
Therefore, the Elvis template is a useful metric for testing more complex
models and is currently the most suitable approach for SED decompositions
involving UV-luminous Type-1 AGNs.

There are two issues in applying the Elvis template. The first is that it is
likely to have a residual contribution in the far infrared from dust heated by
star formation, a possibility that has hindered its application in using the
far infrared to measure star formation rates in quasar host galaxies
\citep[e.g.,][]{ Barnett2015}.  However, a version of the template corrected
for this effect is now available \citep{Xu2015a}.  Based on the analysis of the
Spitzer and IRAS data of the \cite{Elvis1994} sample, these authors found a
tight correlation between the strength of the 11.3 $\mu$m aromatic feature and the
infrared 60 to 25 $\mu$m flux ratio. They concluded that star formation, as
traced by the aromatic feature, boosted the infrared flux ratio in the template
by a factor of 1.27. A scaled \cite{Rieke2009} star-forming galaxy template
(log $L_{\rm IR}$ = 11.0) was subtracted from the \cite{Elvis1994} template to remove
this contribution. The second issue is that of order 10\% of quasars have SEDs
similar to the Elvis template in the UV and optical, but are relatively weak in
the near and mid-infrared, a behavior attributed to a relative lack of hot dust
\citep{Hao2010, Hao2011}. The exact SED shape of these dust-poor quasars
requires future work to address.

\cite{Leipski2013} used three components to represent the AGN SEDs for their
high redshift quasar sample: a UV/optical power-law, a NIR dust emission
component, and a torus model. They adjusted the relative contributions of these
components to optimize their SED fits. However, all three components are
implicitly embedded in the Elvis template. Any adjustments in relative
strengths should only be made after it has been demonstrated that the Elvis
template (or similar ones) gives an unsatisfactory fit. In this work, we use
the Elvis template for our SED decomposition. When combined with the Haro 11
template, we find that its fits are of comparable quality to the relatively
unconstrained fits used by \cite{Leipski2013, Leipski2014} in the sense of
chi-square tests.  There is thus no advantage for our study in using those
more complex and less constrained models for the quasar SEDs - they introduce
additional free parameters without improving the fits correspondingly (see 
Section~\ref{sec:demonstration}).

\subsection{Fitting Procedure}\label{sec:fitting-procedure}

To compare the templates described above to observations, we used a fitting
procedure that takes into account upper-limit data points where available. For
$n$ measurements of $x_i$ with uncertainties $\sigma_i$ and $m$ non-detections
with $x_j<n\sigma_j$ ($n^{\rm th}$ confidence level), we define the fitting
chi-square as
\citep{Isobe1986}:
\begin{equation}
    \chi^2_{\rm total} = \sum\limits_i ^n z_i^2 - \sum\limits_j ^m 2\ln\frac{1+{\rm erf}(z_j/\sqrt{2})}{2} ~,
    \label{eqn:chisq}
\end{equation}
where
\begin{equation}
z_i = \frac{x_i - \hat{x}_i(\theta)}{\sigma_{i}} ~,
\end{equation}
\begin{equation}
    {\rm erf}(x) = \frac{2}{\sqrt{\pi}}\int^{x}_{0}e^{-t^2} dt ~,
\end{equation}
which is the error function, and $\hat{x}_i(\theta)$ is the modeled value. In
Equation~\ref{eqn:chisq}, the first term on the right-hand side is the
classical definition of chi-square, and the second term introduces the error
function to quantify the fitting of upper-limits. We use Markov Chain Monte
Carlo (MCMC) methods to find the parameter set $\theta$ to minimize
$\chi^2_{\rm total}$. 3$\sigma$ upper limits are adopted for all
non-detections. To compare the fitting quality of different fitting methods,
the total $\chi^2_{\rm total}$ should be normalized by the degrees of freedom,
$\nu$. In our case, $\nu = n+m - k$, where $k$ is the number of free parameters
in the model. We will use $\chi_{\nu}^2$ to represent the reduced chi-square,
i.e., $\chi_{\nu}^2 = \chi^2_{\rm total}/\nu$. 

To deal with the trade-off between the goodness of fit and the complexity of
the model, we use the corrected Akaike Information Criterion (AICc) test
\citep{Sugiura1978}, which is defined by
\begin{equation}
{\rm AICc} = -2 \ln \mathcal{L}_{\rm max} + 2k + \frac{2k(k+1)}{N-k-1} ~,
\end{equation}
where $\mathcal{L}_{\rm max}$ is the maximum likelihood achievable by the
model, and $N$ is the number of data points used in the fit, $N=n+m$. The likelihood
of a model to fit data satisfies
\begin{equation}
    -2 \ln \mathcal{L}_{\rm max} = \chi^2_{\rm total} + C ~,
\end{equation}
where the constant $C$ is related to the errors, $\sigma_i$, and the binning,
$\Delta x_i$, of the data points, which are fixed at the time of observations.
We can ignore $C$ when comparing different models to fit the same observations,
and finally have
\begin{equation}
  {\rm AICc} =  \chi^2_{\rm total}  + 2k + \frac{2k(k+1)}{n+m-k-1} ~.
\end{equation}

\section{Tests of the templates}

\subsection{Template Fits for High-$z$ Galaxies}\label{sec:test-galaxy}

\begin{figure}[htp]
    \begin{center} 
	\includegraphics[width=0.8\hsize]{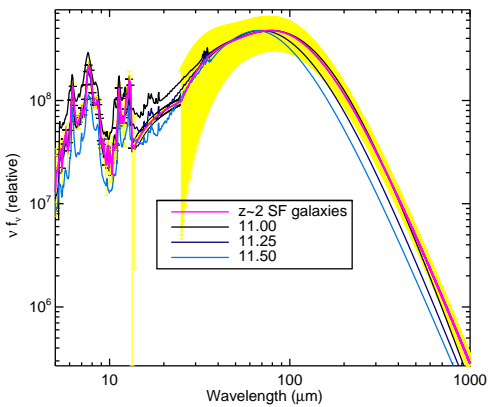} 
	\caption{ 
	    Comparison of the $z\sim2$ galaxy SED in \cite{Kirkpatrick2012}
	    (magenta line with template errors in yellow) and the normal
	    star-forming galaxy templates with $\log L_{\rm IR} = 11.00,
	    11.25, 11.50$ in \cite{Rieke2009} (black, navy, blue lines). All
	    the templates are normalized to have the same $L_{\rm
	    IR,template}$
	}
	\label{fig:R09_template} 
    \end{center} 
\end{figure}

We now discuss alternative SF template candidates to be used at high-$z$.
\cite{Rieke2009} derived templates for local normal star-forming galaxies with
different infrared luminosities ($L_{\rm IR}$).  Although carefully calibrated
in the local Universe, these templates may not apply at high redshift.  The
star formation in luminous galaxies at high-$z$ has been found to be more
physically extended than that in local galaxies with similar $L_{\rm IR}$
(local LIRGs and ULIRGs are sub-kpc, whereas high-$z$ DSFGs are kpc in size,
see \citealt{Rujopakarn2011}).  \cite{Rujopakarn2011,Rujopakarn2013} found that
the \cite{Rieke2009} $\log L_{\rm IR}$=11.00-11.50 SED templates are
representative of galaxies found at $0.4 < z < 2.7$ due to their similar
$L_{\rm IR}$ surface densities. This argument is supported by the consistency
between the empirical average SED of $z\sim2$ galaxies derived in
\cite{Kirkpatrick2012} and the $\log L_{\rm IR}$=11.00-11.50 SED templates from
\cite{Rieke2009} (Figure~\ref{fig:R09_template}). In fact,
Figure~\ref{fig:R09_template} shows the progressively poorer correspondence of
the \cite{Rieke2009} templates with the empirical one with increasing $L_{\rm
IR}$. It is also consistent with the finding of a shift toward colder FIR SEDs
at high redshift by \cite{Symeonidis2009, Symeonidis2013}.  \cite{Greve2012}
found evidence of extended structures in DSFGs out to redshift $z\sim4.0$,
based on analysis of their infrared SEDs.  With this evidence, we focus on
\cite{Rieke2009} SED templates with luminosity $\log L_{\rm IR} < 11.50$ in the
following comparisons.  These normal SF templates represent galaxies that are
almost certainly more metal-rich than is appropriate for $z > 4$. We will
therefore compare them with fits using a template derived from Haro 11.

We test the \cite{Rieke2009} and Haro 11 template fittings to extremely
high-$z$ galaxies, as examples of potential host galaxies for high
redshift quasars. Due to the lack of data available for star-forming galaxies
at $z>5$, we extend our redshift range down to $z=4$. We find 8 galaxies (see
Table~\ref{tab:galaxy-fit}) with multiple constraints on their rest-frame
infrared SEDs, suitable for comparison with these templates. By selection, these highest-$z$
DSFGs are limited to a handful of submillimeter galaxies (SMGs), which were
originally discovered in the submm and relatively bright in the far-IR. The
identification technique of SMGs could bias their SEDs to be relatively cold
compared with high-$z$ galaxies selected in other ways \citep{Le-Floch2004},
whereas the SEDs of low-metallicity galaxies tend to be relatively hot
\citep[e.g.,][]{Remy-Ruyer2013}. As a result, the high-$z$ galaxy examples
studied here might be biased against typical low metallicity galaxies, which,
as in the case of Haro 11, tend to have SEDs dropping rapidly toward the submm.

Table~\ref{tab:galaxy-fit} summarizes the fitting results for the $z>4$
galaxies. We limit the fits to rest-frame 8-1000 $\mum$, where the emission is
purely from dust. Although a few examples, e.g., GN20, have a cold far infrared
SED matched better by the \cite{Rieke2009} templates, in general the fits with
the Haro 11 SED are at least as good. We conclude that it is as good as the
local higher metallicity templates in fitting the SEDs of these extreme $z>4$
SMGs.  That is, even given the selection bias against it, the Haro-11-based
template can be used without a substantial loss of accuracy.

\begin{deluxetable}{lccccc}
    \tabletypesize{\scriptsize}
    \tablewidth{0pt}
    \tablecolumns{6}
    \tablecaption{Comparisons of galaxy templates used to fit $z > 4$ galaxies\label{tab:galaxy-fit}
		}
    \tablehead{
	\colhead{Source} & \colhead{$z$}  & \colhead{$\chi_{\nu,{\rm Haro11}}^2$} & \colhead{R09 best} & \colhead{$\chi_{\nu, {\rm R09}}^2$}  & \colhead{Ref.}\\
	\colhead{(1)} & \colhead{(2)} & \colhead{(3)} & \colhead{(4)} & \colhead{(5)} & \colhead{(6)} 
    }
\startdata
HFLS 3  	&      6.34  &     13.1  & 11.50	& 30.9  &  1\\
AzTEC 3  	&      5.30  &     1.5   &  11.50	& 2.3   &  2, 3\\
HLS J0918+5142  &      5.24  &     18.1  & 11.50	& 18.4  &  3\\
AzTEC 1  	&      4.64  &      6.9  & 11.50	& 2.0  &  3 \\
Capak4.55  	&      4.55  &     0.6   & 11.50 	& 2.1  &  3 \\
ID 141  	&      4.24  &     5.7   &  11.50	& 17.9   &  5\\
GN10 		&      4.05  &     3.2   &  11.50	& 1.5   &  3\\
GN20  		&      4.05  &     23.1  &  11.25	& 0.2   &  3
\enddata
\tablecomments{ 
    Col. (1): Source names sorted by their redshifts; Col. (3): $\chi_{\nu}^2$ of
    Haro 11 template fitting; Col. (4): the \cite{Rieke2009} template which has
    the minimum $\chi_{\nu}^2$; Col. (5): minimum $\chi_{\nu}^2$ among tested
    \cite{Rieke2009} templates; Col. (6): references for photometric data.\\
    {\bf References.} (1) \cite{Riechers2013}; (2) \cite{Dwek2011}; (3)
\cite{Huang2014};  (4) \cite{Rawle2014};  (5) \cite{Cox2011} }
\end{deluxetable}

\subsection{Template Fits for High-$z$ Quasars}

\begin{figure}[htp] 
    \begin{center} 
	\includegraphics[width=0.9\hsize]{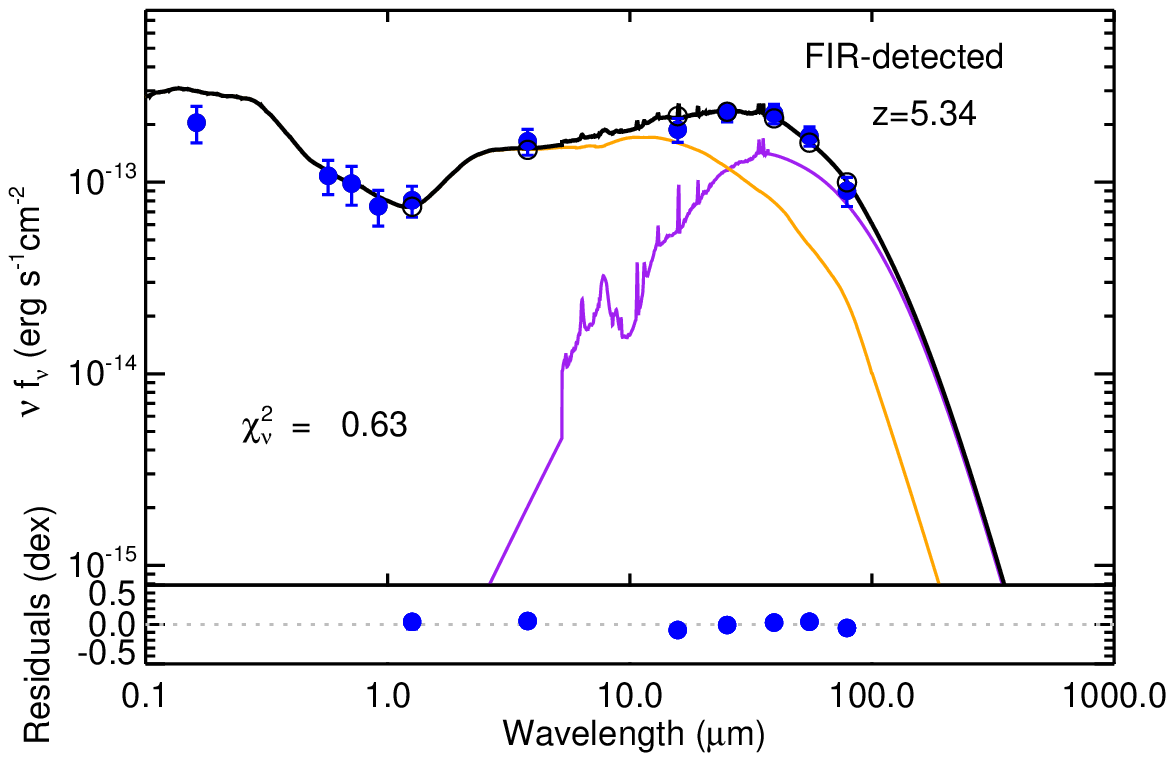} 
	\includegraphics[width=0.9\hsize]{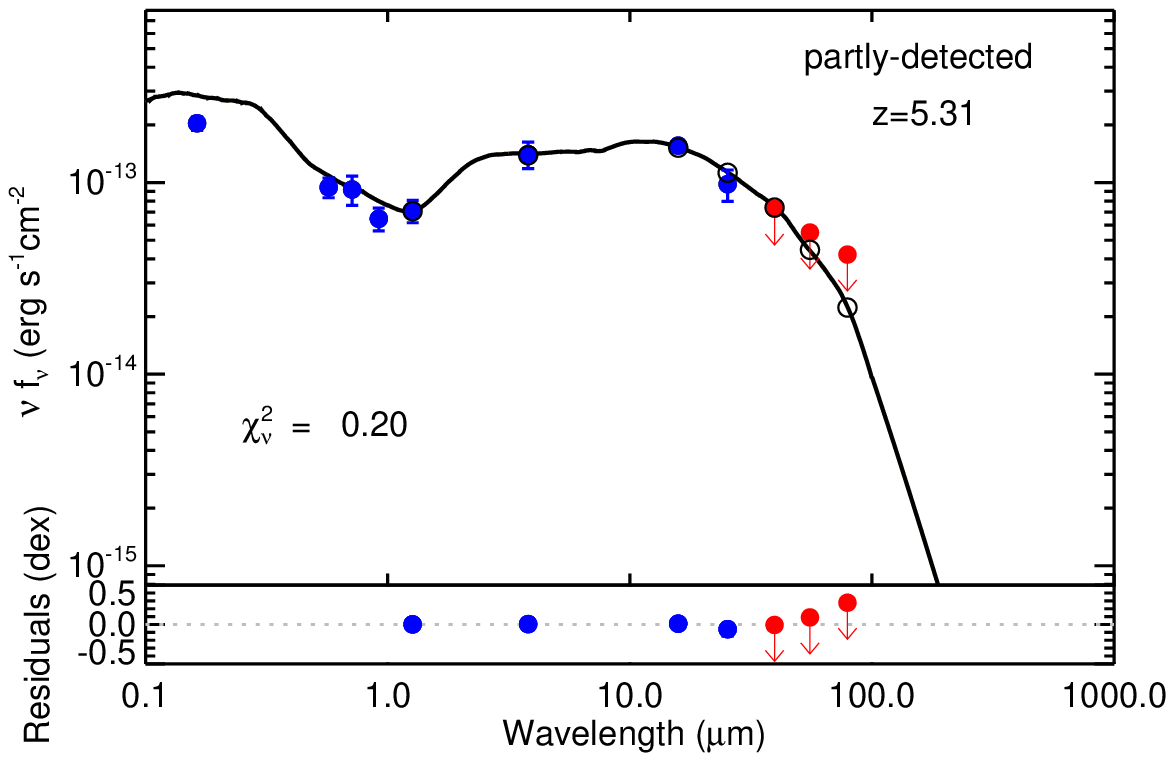} 
	\includegraphics[width=0.9\hsize]{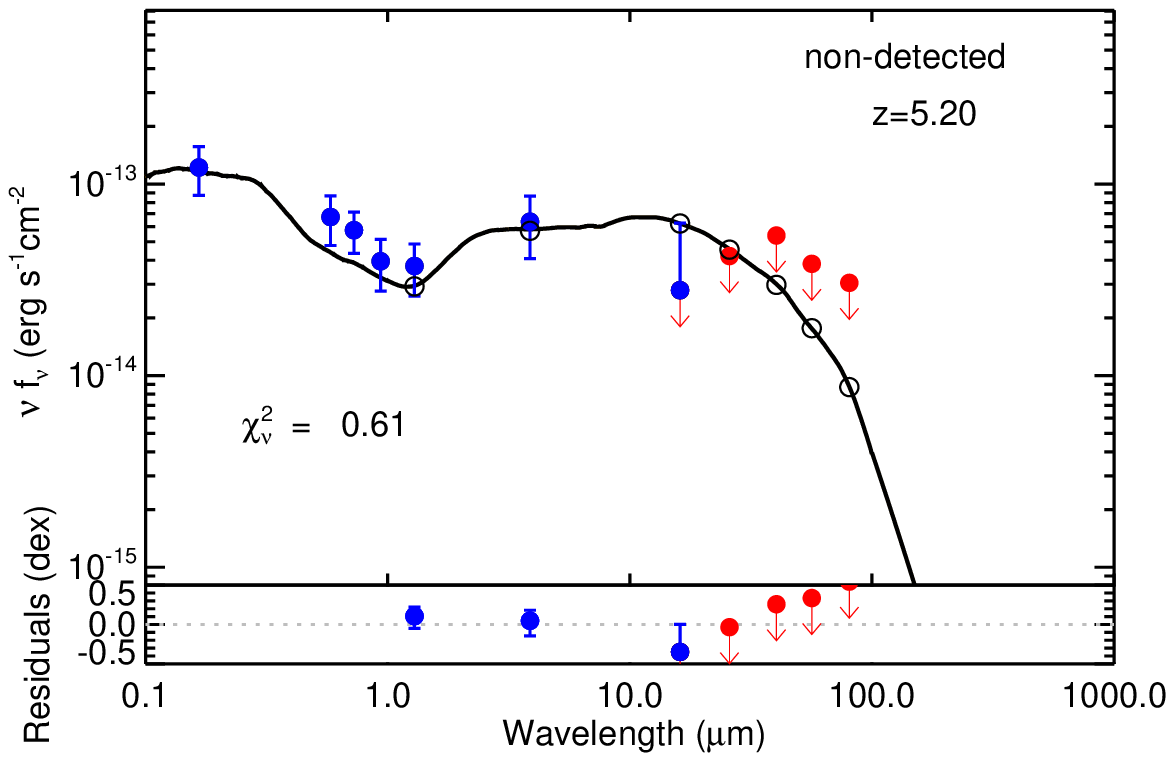} 
	\caption{
	    Three stacked SEDs in \cite{Leipski2014} and our fitting results.
	    We limit the fitting to rest-frame wavelengths 1-1000$\mum$.  The
	    black solid line is the modelled SED, the black circles are the
	    modelled points (convolved with the corresponding photometric
	    filters).  The AGN component and galaxy component are shown as gray
	    and purple solid lines, respectively. The blue dots are for
	    detections with 1$\sigma$ error bars and red dots are for upper
	    limits at 3$\sigma$. The reduced chi-square of the fit is shown on
	    the bottom left of each plot.
	} 
	\label{fig:stacked_sed} 
    \end{center} 
\end{figure}

The \cite{Elvis1994} template has been shown to match type-1 quasar SEDs for
redshifts up to $z\sim3$ and for wavelengths $\lambda \lesssim24\mum$
(\citealt{Hao2014, Xu2015a}). An issue in applying it, or the similar template
of \cite{Richards2006}, in the far-IR is the uncertain contribution of host
galaxy star formation \citep[e.g.,][]{Barnett2015}.  However, \cite{Xu2015a}
were able to correct for this effect. In Figure~\ref{fig:stacked_sed}, we
compare this corrected template with the stacked SEDs from \cite{Leipski2014}.
While the UV-to-optical parts of all three SEDs are well matched with the AGN
continuum template, differences emerge in the infrared.  The stacked SED of
quasars detected in at least 3 {\it Herschel} bands has a substantial excess
over the AGN template in the far-IR, which we attribute to host galaxy star
formation (see Section~\ref{sec:heating}). The stacked SED of quasars not
detected with \herschel is not matched as well in the infrared although the
reduced chi-square is still acceptable. This behavior could be due to the
unsuitability of a classical AGN template to represent the hot-dust-free
\citep{Jiang2010} or hot-dust-poor \citep{Hao2010, Hao2011} quasars (hereafter
hot-dust-deficient quasars, or HDD quasars) as pointed out by
\cite{Leipski2014}. The fit to the {\it Herschel} partly-detected (detected in
only 1-2 {\it Herschel} bands) stacked SED is virtually perfect over the entire
wavelength range. The agreement of the template with both the {\it
Herschel}-undetected and {\it Herschel} partly-detected stacked SEDs suggests
that the star-formation corrected Elvis SED is a good choice to fit the
high-redshift AGN continua. More discussion will be provided in
Section~\ref{sec:discussion}

\subsection{SED Fitting with Well-measured High-$z$ Quasars}\label{sec:demonstration}

To test further whether the Haro 11 template as well as the AGN
(modified \cite{Elvis1994}) template are reasonable choices to decompose $z
\gtrsim 5$ quasar SEDs, we focus on 5 quasars with the most complete infrared
SEDs. Besides SDSS J1204$-$0021, the SEDs of all the other quasars were studied
in \cite{Leipski2013}\footnote{Among the five millimeter-detected quasars with
    at least two \herschel observations in \cite{Leipski2013}, SDSS
    J1044$-$0125 is excluded since the number of detected data points at
    rest-frame $1-1000\mum$ is smaller than the number of free parameters of
    the \cite{Leipski2013} model. This hinders the computation of a reduced
chi-square of the \cite{Leipski2013} model for only the IR data points to be
compared with the two-component fits.}. We model the observed rest-frame
1-1000$\mum$ SED as a linear combination of the Haro 11 template and the AGN
template with two free normalizing factors.  These two templates are taken to
be independent.  To compare the Haro 11 template with the normal SF templates,
we replace the Haro 11 template by the normal SF templates in \cite{Rieke2009},
and redo the fit. We also apply the \cite{Leipski2013} model to the UV-to-IR
SED of these quasars and compare the fits of the IR SED with those from our
two-component models.

In Figure~\ref{fig:qso-decomp-com-1}, we present the SED decomposition results.
In general, the Haro 11 template fits have smaller residuals ($\lesssim$
0.3 dex) compared with the best-fit\footnote{We fit these quasars with normal
    SF templates in \cite{Rieke2009} with $\log L_{\rm IR}\le 11.50$, and pick
    the one that has the lowest $\chi_{\nu}^2$ as the best. The $\log L_{\rm IR} > 11.50$
normal SF templates do not yield any better results, especially in the mid-IR,
as shown in Figure~\ref{fig:qso-decomp-com-1}.} normal SF template.  In
particular, the Haro 11 template yields much better fits in representing
the warm dust component from the two-component decomposition. We comment on the
two-component fits (left and middle columns of
Figure~\ref{fig:qso-decomp-com-1}) for each SED below:

{\bf J0338+0021} (or SDSS J033829.31+002156.3; we use JHHMM $\pm$ DDMM for
brevity). The Haro 11 template is better than the best normal SF template
in decomposing the SED.  Fitting the mid-IR at $< 10\mum$ and the far-IR drop
beyond $100 \mum$ results in the normal SF template model underestimating the
flux at $10-100 \mum$.  We note an excess between $\sim$ 10 $\mum$ and $\sim40$
$\mum$ over the normal SF template fitting model SED, which could be the warm excess
seen in relatively low-$z$ AGN SEDs reported by \cite{Xu2015a}. In contrast, such
an excess is not strong in the Haro 11 template fits.

{\bf J0756+4104}. Judging from the fit $\chi_\nu^2$, the normal SF template seems
better. However, $\sim50\%$ of the $\chi_{\nu}^2$ of the Haro 11 template fit is
contributed by the data point at the longest wavelength ($\lambda_{\rm rest} =
139\mum$), whereas the $\chi_{\nu}^2$ contribution of the same data point in the
normal SF template fitting is minimal. Again, the normal SF template fitting
underestimates the SED at $\sim 10-40\mum$. We conclude the normal SF template
and the Haro 11 template yield fits of similar quality.

{\bf J0927+2001}. The Haro 11 template is much better than the normal SF
template in reproducing the observed SED. The maximum deviation of the dwarf
galaxy model and observed SEDs is less than 0.3 dex. In the case of this
quasar, the normal SF template underestimates the SED at $\sim 10-100\mum$. 

{\bf J1148+5251}. For this well-studied quasar, the Haro 11 template
fitting is almost the same as the best normal SF template fitting when
comparing $\chi_{\nu}^2$. Interestingly, our estimation of the host contribution of
this quasar is consistent with result based on the theoretical analysis by
\cite{Schneider2014} .

{\bf J1204$-$0021}. This is the only case where the Haro 11 template fitting
has one data point with fitting residual (slightly) greater than 0.3 dex. Both
two-component fits underestimate the observed $10-100\mum$ flux. However, the
residual from the Haro 11 template fitting is much smaller than the normal SF
one.

~

For the \cite{Leipski2013} model (right column of
Figure~\ref{fig:qso-decomp-com-1}), we only apply the fit to the detected data
points in the UV-to-IR, in the same fashion as \cite{Leipski2013}, and compute
the $\chi_{\nu}^2$ for the detected data points at rest-frame 1-1000$\mum$.
Since it has more components, especially a torus component selected from a
large model library, small details of the observed SED can be reproduced. Thus,
the residuals are generally smaller. However, our two-parameter fit has similar
reduced chi-square compared with the \cite{Leipski2013} model, despite its
simplicity. To judge which fit is preferred, we have used the AICc test (see
Section~\ref{sec:fitting-procedure}).  Since the slope of the power law
component is not useful in fitting the infrared data, we have assumed that the
\cite{Leipski2013} fits had six free parameters over 1-1000 $\mum$. As shown in
Table~\ref{tab:qso-fit}, the value of AICc is lower in all five cases for the
2-parameter fits, indicating that they are indeed preferred. That is, even for
these quasars with the maximum number of measurements, the \cite{Leipski2013}
model over-fits the data compared with our two-parameter one.

\begin{figure*}[htbp] 
    \centering 
    \begin{tabular}{@{}ccc@{}}
    \includegraphics[width=0.3\hsize]{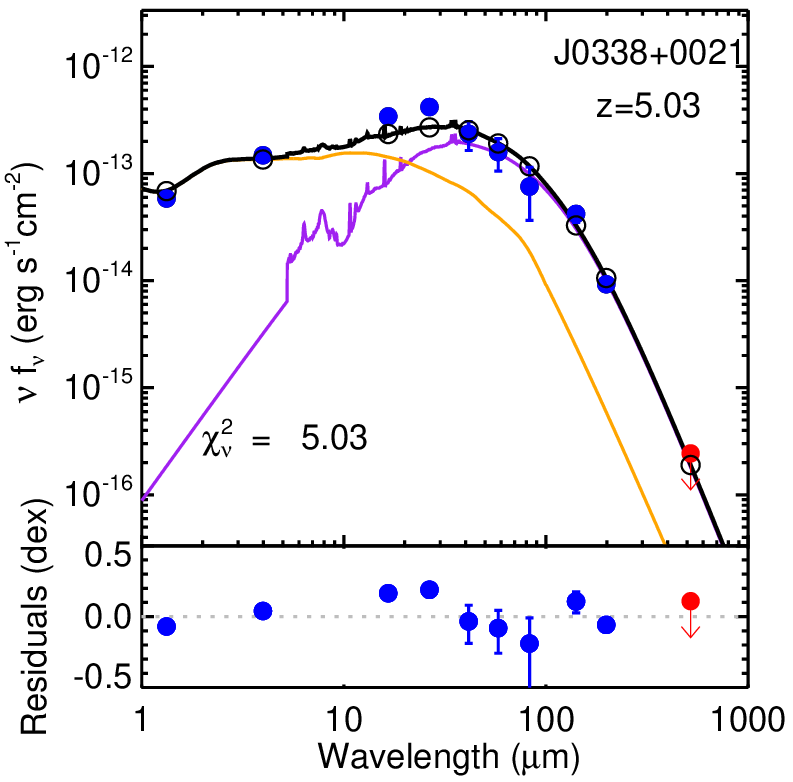} &
    \includegraphics[width=0.3\hsize]{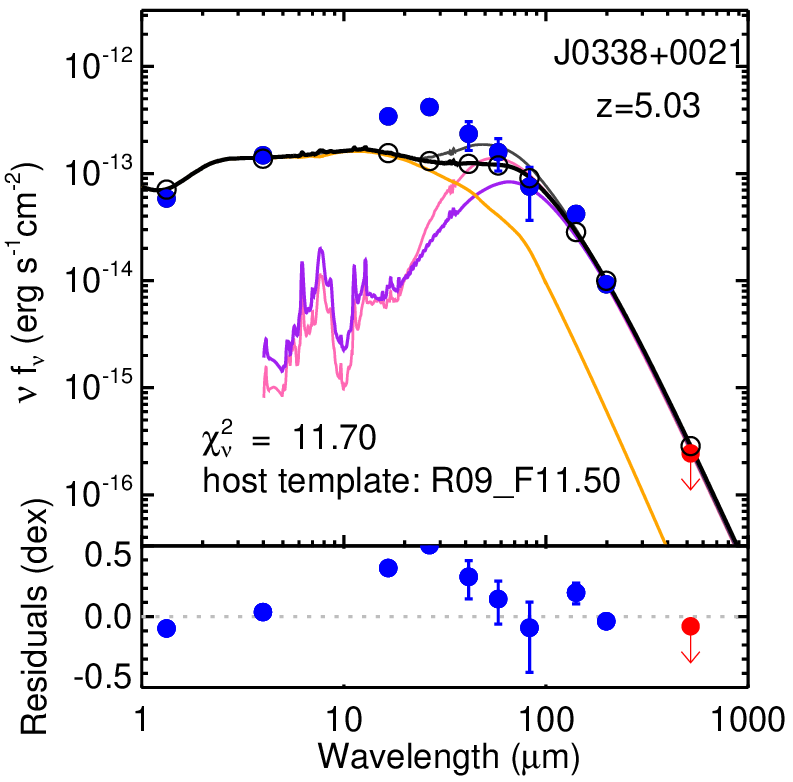} &
    \includegraphics[width=0.3\hsize]{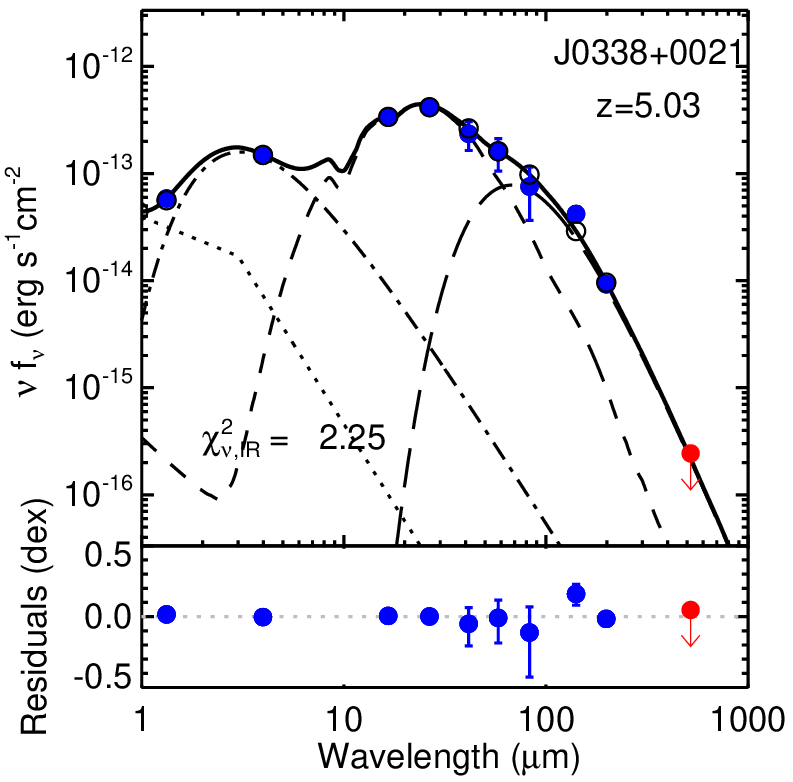} \\
    \includegraphics[width=0.3\hsize]{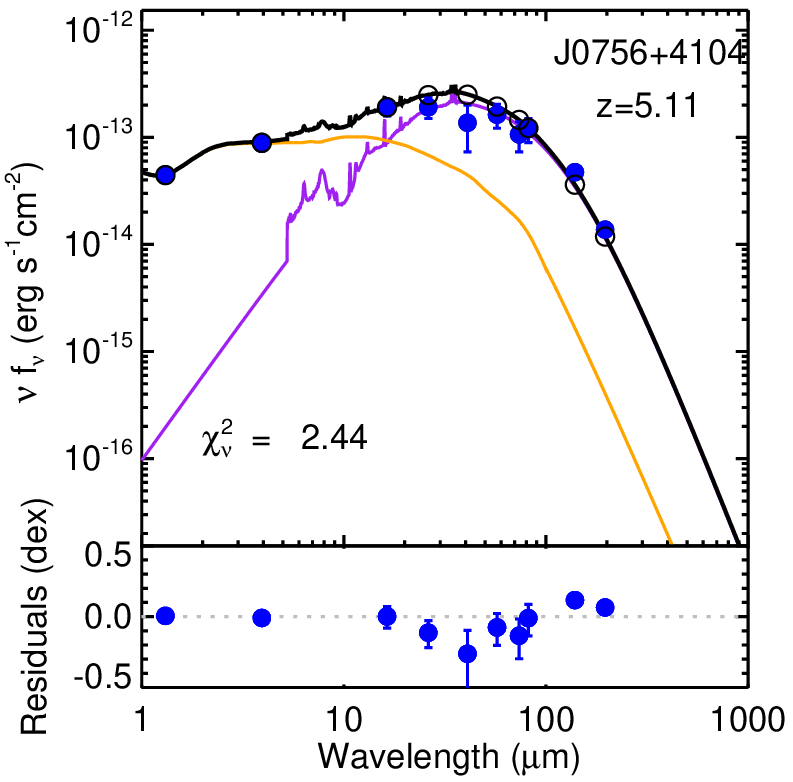} &
    \includegraphics[width=0.3\hsize]{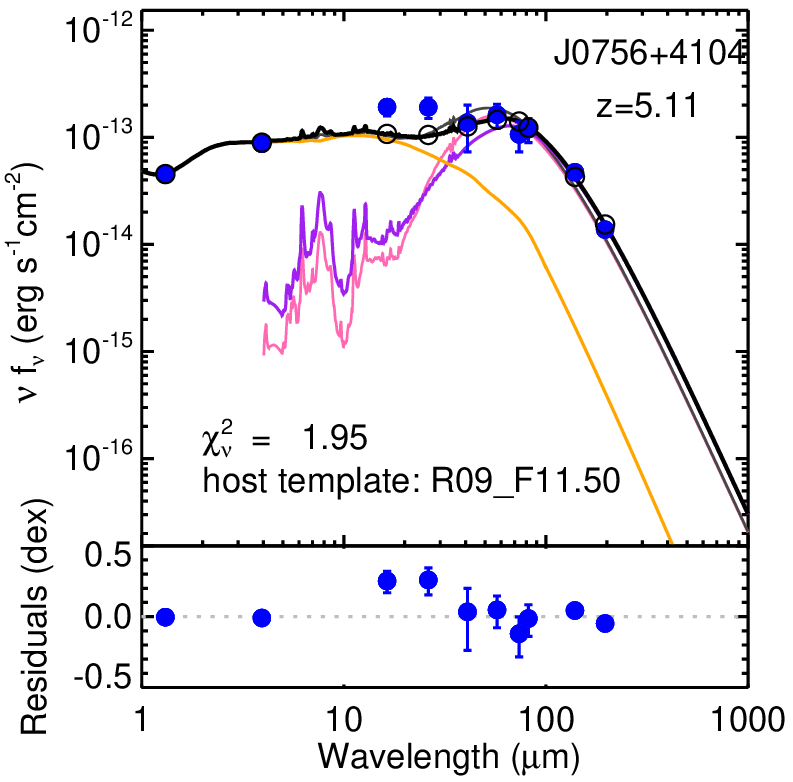} &
    \includegraphics[width=0.3\hsize]{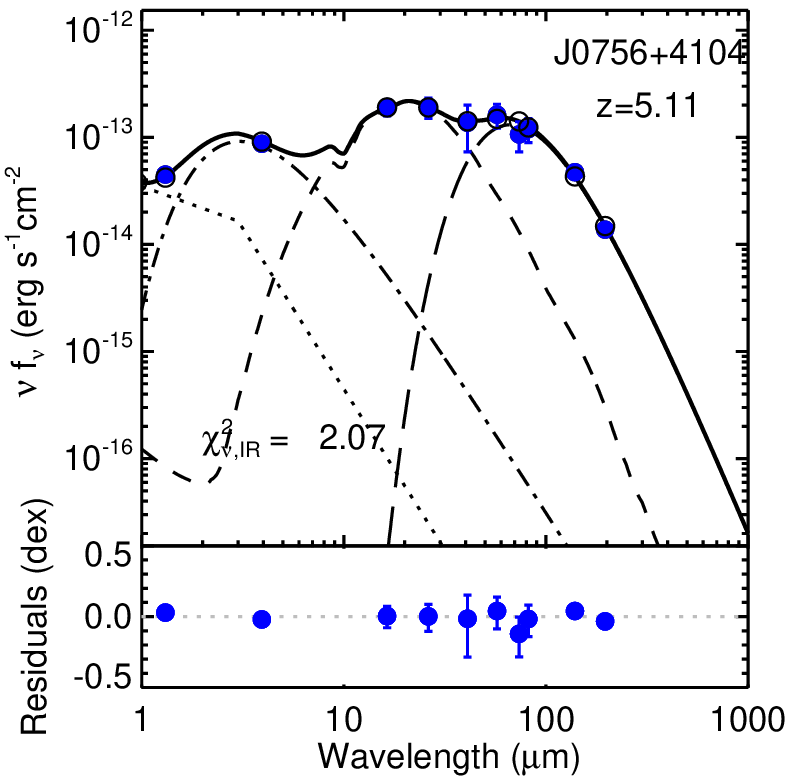} \\
    \includegraphics[width=0.3\hsize]{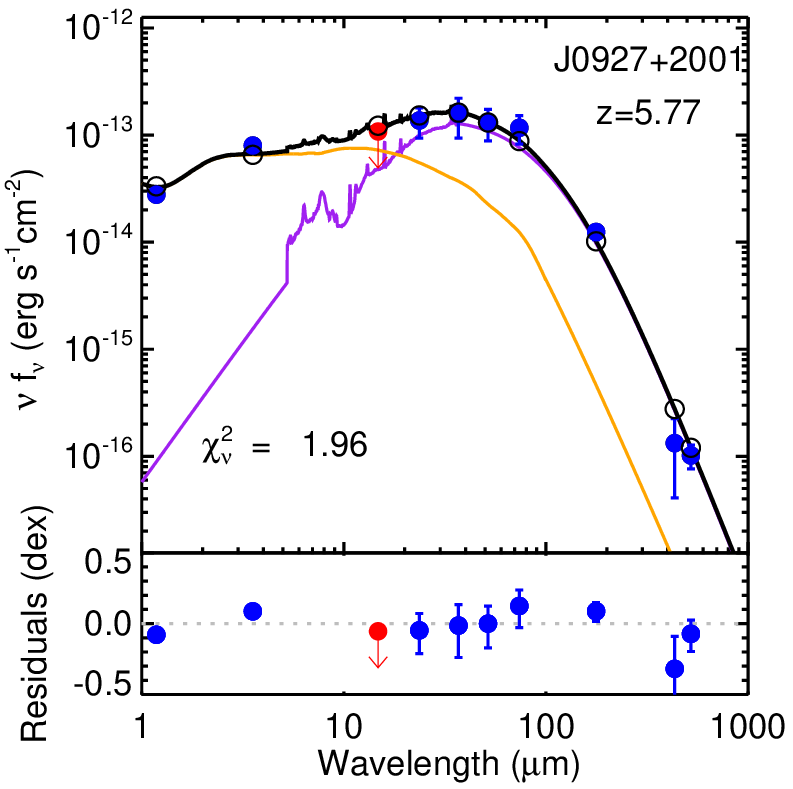} &
    \includegraphics[width=0.3\hsize]{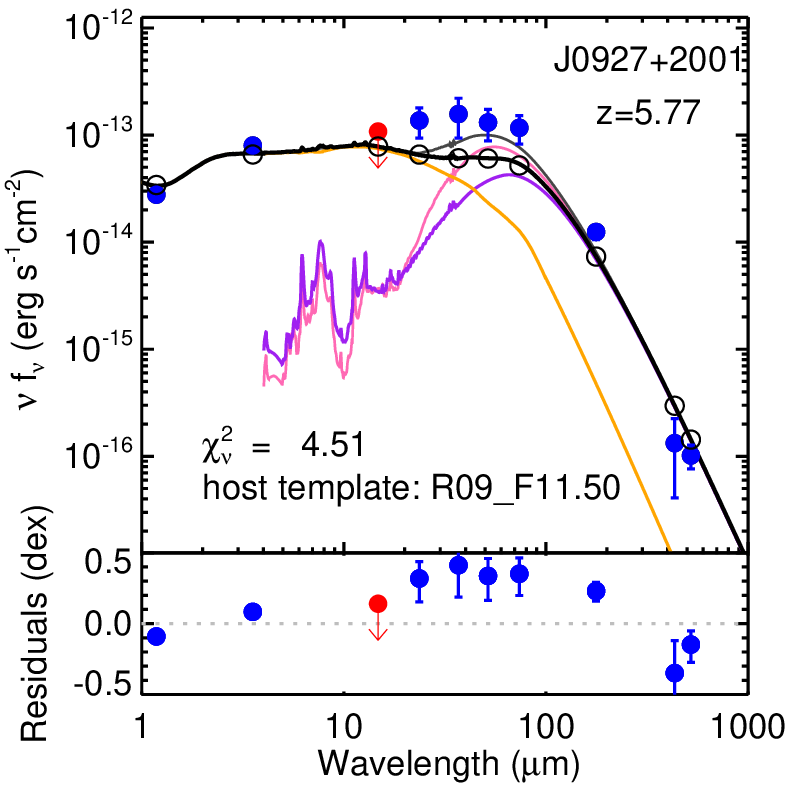} &
    \includegraphics[width=0.3\hsize]{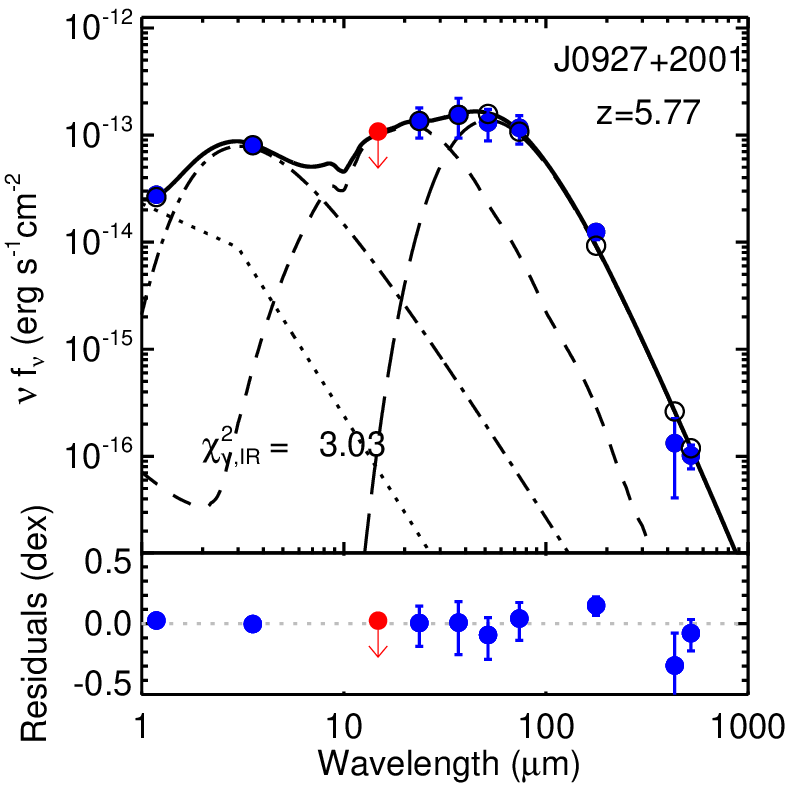} 
    \end{tabular} 
\caption{\label{fig:qso-decomp-com-0}
    SEDs and their decompositions for 5 quasars with strong far-IR SED
    constraints. The plots show $\nu F_{\nu}$ in units of ${\rm erg/s/cm^2}$ over the
    rest-frame wavelengths. We show the results from ``Haro 11 + AGN''
    decomposition (left column), ``normal SF galaxy + AGN'' decomposition
    (middle column), and the \cite{Leipski2013} model (right column). In each
    plot, the blue dots are for detections with 1$\sigma$ error bars and red
    dots are for upper limits at 3$\sigma$. The bottom panel of each plot
    presents the residuals from the fit.  In the left and middle columns, the
    colored lines show the results of a two-component SED fit as described in
    Section~\ref{sec:templates}: the SF-subtracted type-1 AGN template in
    orange, and a galaxy template in purple. The black line is the total of
    these two components. The black circles are the synthetic photometry points
    from the model. To demonstrate that \cite{Rieke2009} templates with $\log
    L_{\rm IR} > 11.50$ do not yield better results, we make another
    fitting with \cite{Rieke2009} $\log L_{\rm IR}$=12.50 template, and plot
    the fitted galaxy component (pink thin line) and the total model SED (gray
    thin line) in the middle column.  In the right column, we use the
    \cite{Leipski2013} model, which is a combination of a UV/optical power-law
    (dotted line), a 1200 K near-IR dust component (dot-dashed line), a
    near-/mid-IR torus model (short-dashed line), and a modified black-body
    far-IR component with $\beta=1.6$ (long-dashed line). 
} 
\end{figure*}

\addtocounter{figure}{-1}
\begin{figure*}[htbp] 
    \centering 
    \begin{tabular}{@{}ccc@{}}
    \includegraphics[width=0.3\hsize]{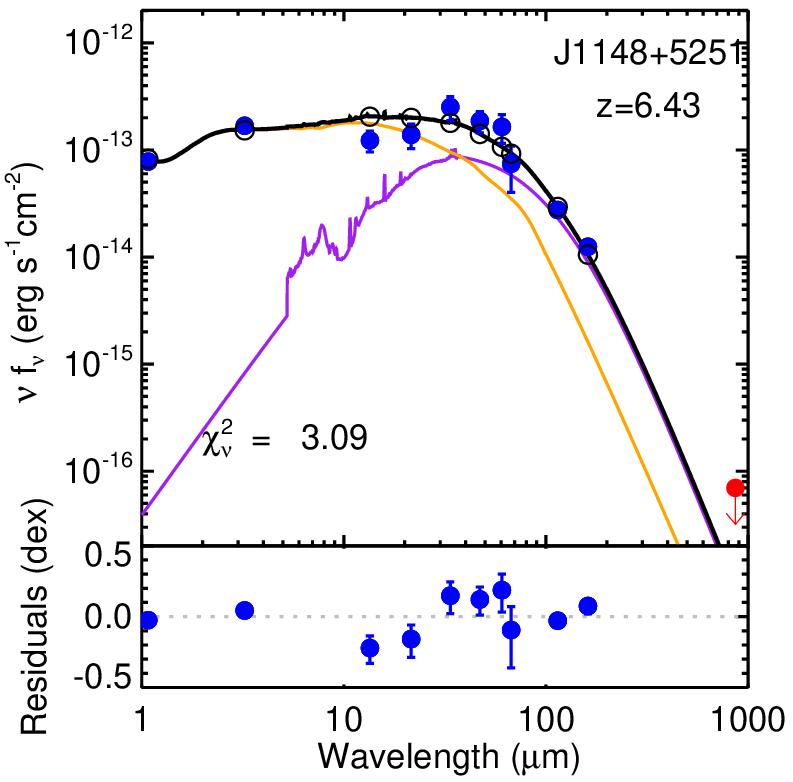} &
    \includegraphics[width=0.3\hsize]{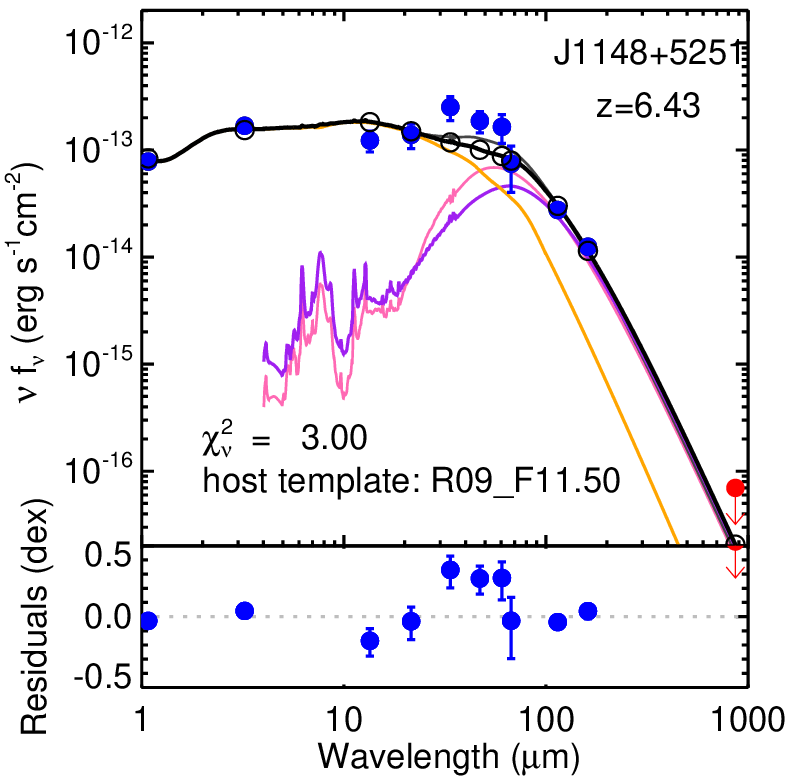} &
    \includegraphics[width=0.3\hsize]{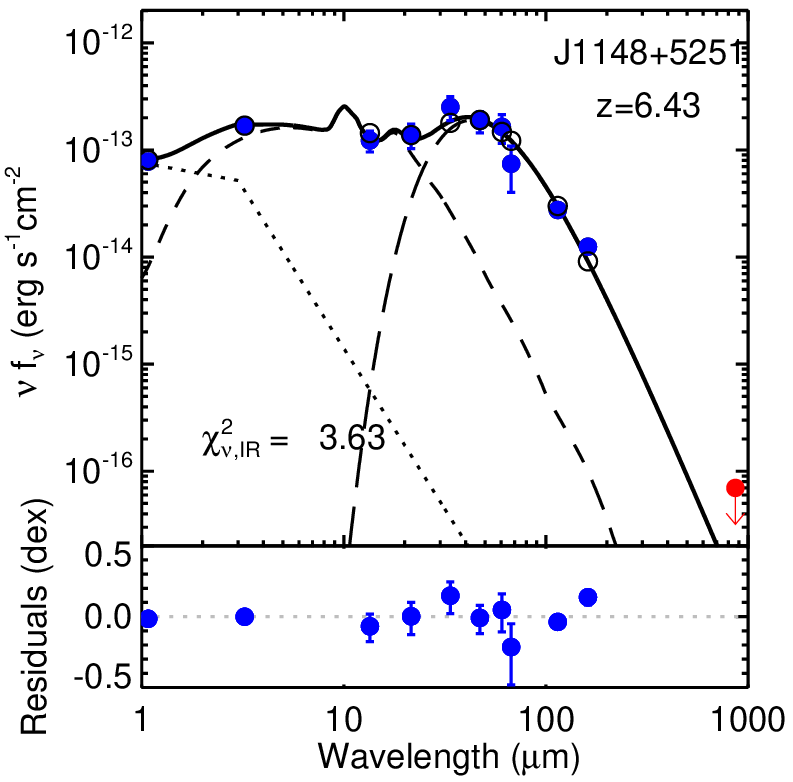} \\
    \includegraphics[width=0.3\hsize]{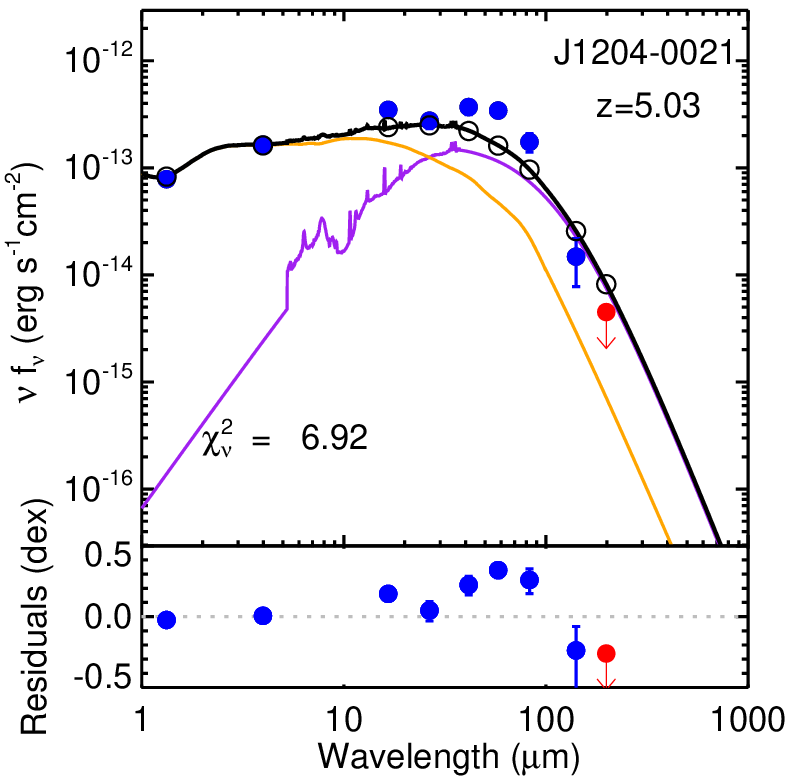} &
    \includegraphics[width=0.3\hsize]{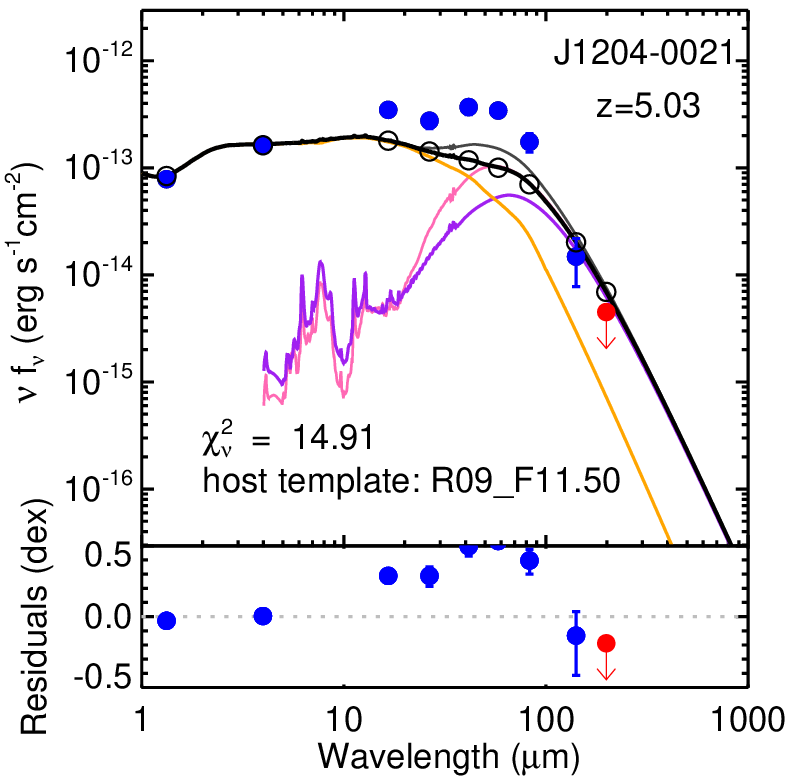} &
    \includegraphics[width=0.3\hsize]{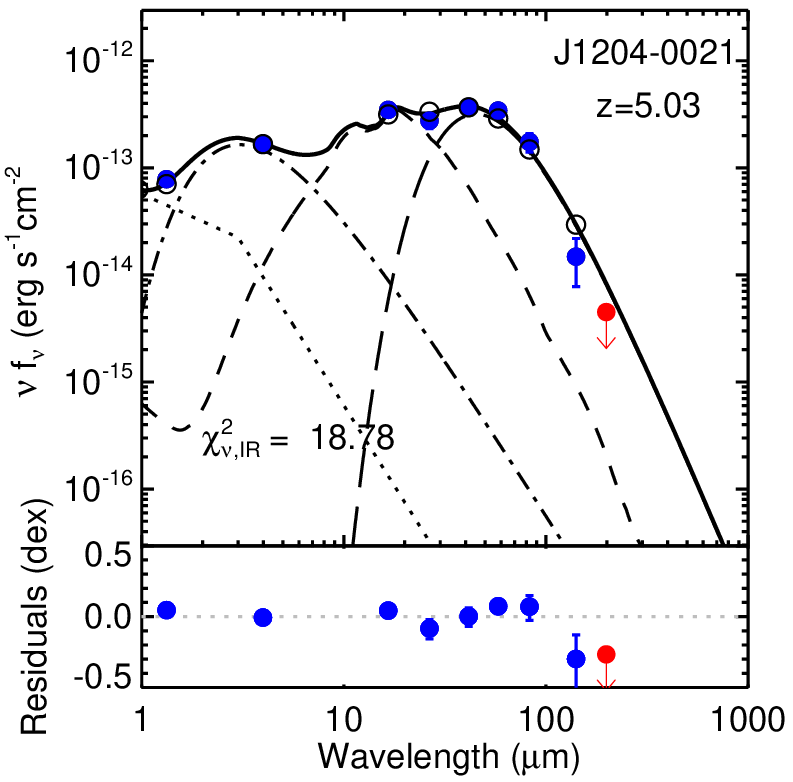} 
    \end{tabular} 
\caption{\label{fig:qso-decomp-com-1}
    continued. SEDs and their decompositions of 5 quasars with strong far-IR SED
    constraints.} 
\end{figure*}

\begin{deluxetable*}{lccccc|cc|cc}
    \tablewidth{0pt}
    \tablecolumns{10}
    \tablecaption{SED decomposition results for $z\gtrsim5$ 5 quasars with well-measured SEDs\label{tab:qso-fit}
		}
    \tablehead{
	\colhead{Source} & \colhead{Redshift} & \colhead{ $L_{\rm IR}$/$10^{13}L_{\odot}$} & \colhead{$f_{\rm host, IR}$} & \colhead{$\chi_{\nu}^2$} & \colhead{AICc} & \colhead{R09 best} & \colhead{$\chi_{\nu, {\rm R09}}^2$} & \colhead{$\chi^2_{\nu, {\rm L13,IR}}$}  & \colhead{AICc$_{\rm L13}$} \\
	\colhead{(1)} & \colhead{(2)} & \colhead{(3)} & \colhead{(4)} & \colhead{(5)} & \colhead{(6)}  & \colhead{(7)} & \colhead{(8)}  & \colhead{(9)}  & \colhead{(10)}
     }
\startdata
SDSS J0338+0021	   &      5.03  & 4.01 &    0.57  &  5.03  & 45.95   & 11.50	  & 11.70  & 2.25   & 49.00  \\
SDSS J0756+4104    &      5.11  & 3.77 &    0.69  &  2.44  & 25.26   & 11.50      &  1.95  & 2.07   & 48.28  \\
SDSS J0927+2001    &      5.77  & 3.21 &    0.64  &  1.96  & 21.39   & 11.50      &  4.51  & 3.03   & 52.12  \\
SDSS J1148+5251    &      6.43  & 5.31 &    0.34  &  3.09  & 33.31   & 11.50      &  3.00  & 3.63   & 51.10 \\
SDSS J1204$-$0021  &      5.03  & 3.79 &    0.46  &  6.92  & 54.44   & 11.50      & 14.91  & 18.78  & 110.34
\enddata
\tablecomments{
    Results of full IR fits. Upper-limit data points are included in the
    evaluation process.  Col. (3): the total infrared luminosity (8-1000
    $\mum$) estimated from the ``Haro 11 + AGN'' two-component SED fit; Col.
    (4): the fraction of luminosity of host template contribution to the whole
    fit SED, based on result from the ``Haro 11 + AGN'' decomposition; Col.
    (5): reduced chi-square from the ``Haro 11 + AGN'' decomposition; Col.
    (6): the AICc test value of the ``Haro 11 + AGN'' two-component model; Col.
    (7): log$L_{\rm IR}$ of the \cite{Rieke2009} template which has the minimum
    $\chi_\nu^2$; Col. (8): the minimum $\chi_\nu^2$ of the ``normal SF galaxy + AGN''
    decomposition with all tested \cite{Rieke2009} templates; Col. (9): reduced
    chi-square from the \cite{Leipski2013} model, only counting
    data points at rest-frame 1-1000 $\mum$; Col. (10): the AICc test
    value of the \cite{Leipski2013} model, assuming 6 free parameters.
}
\end{deluxetable*}

In summary, we find that the Haro 11 galaxy template and the modified Elvis AGN
template are at least as good at fitting the overall high-$z$ quasar SEDs as
the fits using templates for local star forming galaxies of solar metallicity.
The Haro 11 template fits better in the rest-frame mid-IR, but may be slightly
worse in the far-IR range. Though the \cite{Leipski2013} model could reproduce
more details of the observed SED, our two-component model yields fits of
comparable overall quality and is preferred in model selection due to its
simplicity.

\section{AGN and host galaxy decomposition for quasars at $z \gtrsim 5$} \label{sec:sfr-qso}

Combining {\it Herschel}, \spitzer and ground-based 250 GHz observations, we
apply the ``Haro 11 + AGN'' two-component fit to the observed SEDs of all
69 quasars in \cite{Leipski2014} (hereafter, {\it sample-A}). Since our
interest is the infrared output, we again limit the fit to rest-frame
$1.0-1000\mum$.  The fits are presented in Figure~\ref{fig:qso_all1}. We
can confirm again the suitability of the modified \cite{Elvis1994} AGN
template: the near-IR to mid-IR SEDs of these $z\gtrsim5$ quasars are
well-matched with our empirical AGN template in 58 cases. Combining with the
Haro 11 template, this modified \cite{Elvis1994} AGN template provides
reasonably good fittings to all these $z\gtrsim5$ quasars, which suggests the
\cite{Elvis1994} AGN template derived on local quasars is suitable for vast
majority of $z\gtrsim5$ quasars. There are 11 cases where the Elvis template
overestimates the mid-infrared, which is the signature of HDD quasars: the
incidence of this behavior ($\sim16\%$) is similar to that observed at lower
redshift \citep{Mor2011}. The $\chi_{\nu}^2$, derived total infrared luminosities
($L({\rm IR})$), and host component contributions from the fits are listed in
Table~\ref{tab:qso-all-fit}. We also calculate the star formation rates using
the method described in Section~\ref{sec:qso-sfr}

\begin{figure*}[htp]
    \begin{center} 
	\includegraphics[width=1.0\hsize]{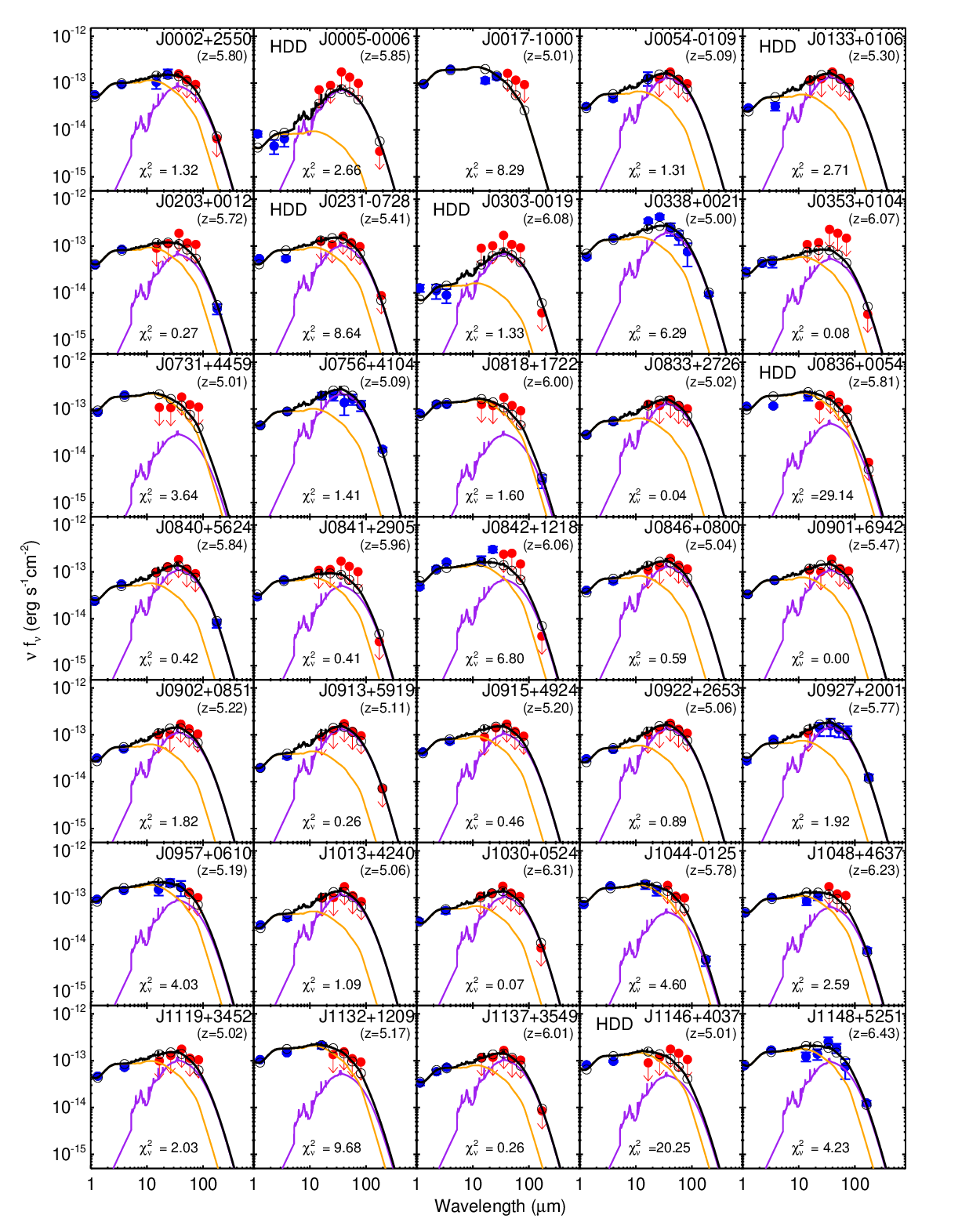} 
	\caption{ 
	    Two-component SED decompositions of 69 quasars ({\it sample-A}) in
	    \cite{Leipski2014}.  The plots show $\nu F_{\nu}$ in unit of
	    ${\rm erg/s/cm^2}$ over the rest-frame wavelength. The black solid line is
	    the modelled SED, the black circles are the modelled points. The
	    AGN component and galaxy component are shown as gray and
	    purple solid lines, respectively. The blue dots are for detections
	    with 1$\sigma$ error bars and red dots are for upper limits at
	    3$\sigma$. The reduced chi-square of the fit is shown on the bottom
	    of each plot. We put a tag `HDD' near the top-left corner of the panel
	    if the corresponding quasar is identified as hot-dust-deficient (HDD).
	}
	\label{fig:qso_all1} 
    \end{center} 
\end{figure*}

\addtocounter{figure}{-1}
\begin{figure*}[htp]
    \begin{center} 
	\includegraphics[width=1.0\hsize]{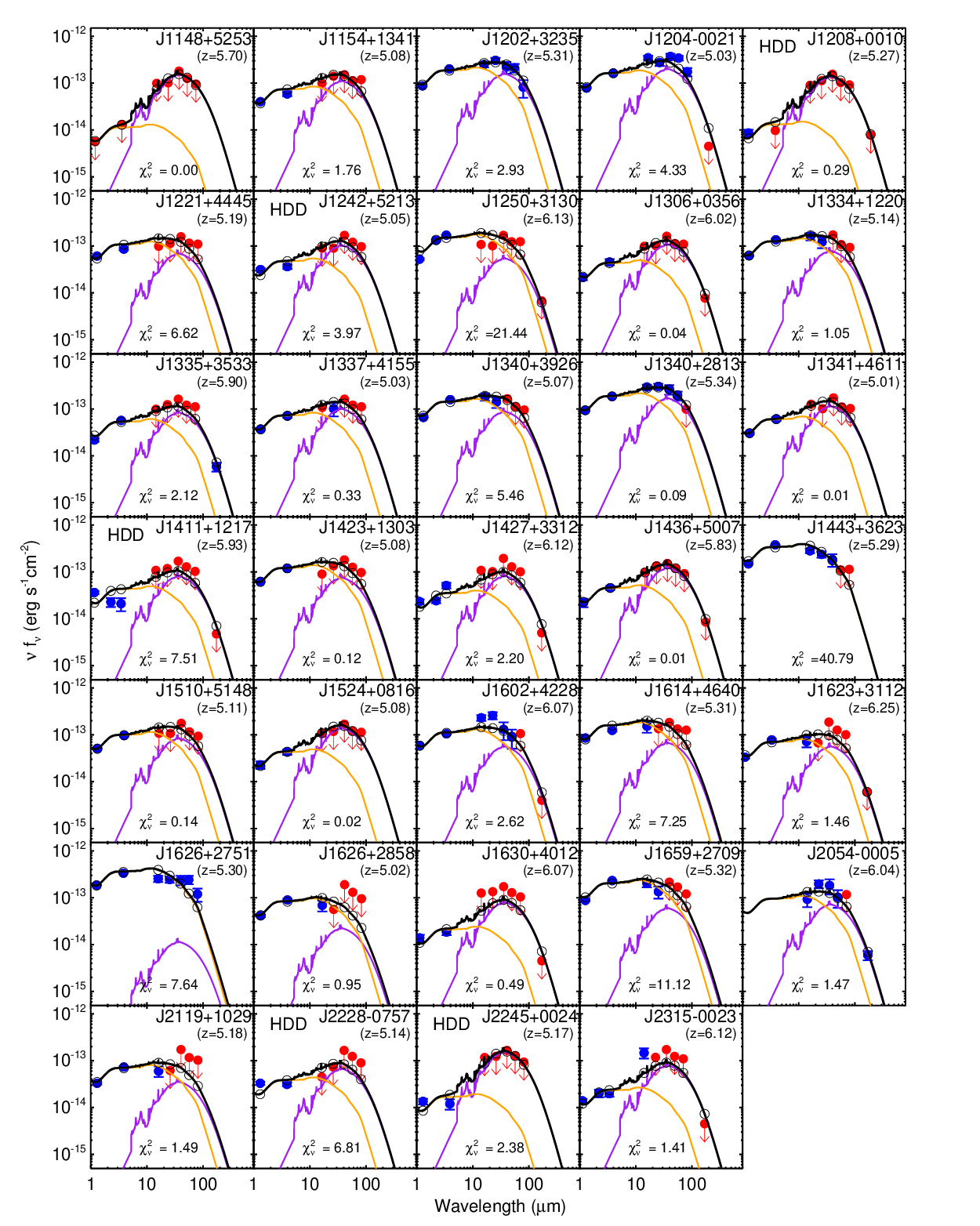} 
	\caption{ 
	    continued. Two-component SED decomposition of 69 quasars ({\it
	    sample-A}) in \cite{Leipski2014}.
	}
	\label{fig:qso_all2} 
    \end{center} 
\end{figure*}

    We can compare the results of the host galaxy far-IR luminosity with other
    works. It is frequently assumed that the far-IR SED of high-$z$
    quasars can be described as a $T=47~$K and $\beta =1.6$ modified
    black body \citep{Beelen2006} and the infrared luminosity of
    this component is used to estimate their star formation rates
    \citep[e.g.,][]{Omont2013,Leipski2013,Leipski2014,Calura2014,Willott2015}.
    However, real galaxies have strong mid-infrared emission that is
    under-represented by a single (modified) black body SED
    \citep[e.g.,][]{Dunne2001,Willmer2009, Galametz2012,Kirkpatrick2012}.  With
    the inclusion of the mid-IR energy contribution from star formation, the
    Haro 11 template yields an infrared luminosity $L_{\rm SF, IR}$ 1.5-2.0
    times larger than the infrared luminosity of the \cite{Beelen2006} modified
    back body template $L_{\rm FIR}$, depending on the exact far-IR constraints
    to the galaxy component.  Compared with the \cite{Leipski2014} results from
    the four-component decomposition on 19 quasars with most complete IR SED
    observations, our results are different, especially for the host galaxy
    far-IR luminosity.  For a quasar with strong host galaxy far-IR emission,
    its mid-IR emission will be dominated by star formation. Since the modified
    black body misses significant luminosity at shorter wavelengths, the
    \cite{Leipski2013} model has to scale up the torus component to fit the
    rest-frame mid-IR data, ending with an overestimated fraction of the far-IR
    emission due to the AGN (the torus) and an underestimated host galaxy
    contribution. Consequently, the star formation rates derived by the
    \cite{Leipski2013} model in such cases (like SDSS J0756+4104, J0927+2001,
    J1202+3235, J1340+2813, which have $L_{\rm SF,IR} \sim 2.2-2.9L_{\rm FIR}$)
    are much lower than our values.  For a quasar without strong host galaxy
    far-IR emission, however, the \cite{Leipski2013} model gives a higher
    $L_{\rm FIR}$ than our $L_{\rm SF,IR}$. Examples are SDSS J0842+1218,
    J1044-0125, J1048+4637, J1148+5251, J1659+2709, with $L_{\rm SF,IR} \sim
    0.5-1.0 L_{\rm FIR}$. This discrepancy is still due to the template
    differences. The torus templates used in \cite{Leipski2013,Leipski2014} by
    themselves generally underpredict the far-IR emission compared with our AGN
    template. As a result, the far-IR modified black body has to be scaled up
    to match the far-IR observations.  We believe the star formation rate based
    on the $L_{\rm SF,IR}$ from our model is more reliable, since 1) the host
    galaxy template is based on a real galaxy that includes the mid-IR star
    formation contribution; 2) the relative contributions of the torus and
    near-IR component are fixed in our AGN template that is based on real
    observations. In addition, thanks to the simplicity of the two-component
    model, we can place constraints on the host galaxy star formation for the
    other 50 $z\gtrsim5$ quasars in \cite{Leipski2014}, which only have
    upper-limits for the MIR-FIR SEDs and the \cite{Leipski2013} model can not
    fit. We will discuss our model results in Section~\ref{sec:discussion}.

\begin{deluxetable*}{lccccccc}
    \tabletypesize{\scriptsize}
    \tablecaption{SED decomposition results for $z\gtrsim5.0$ 69 quasars ({\it sample-A})\label{tab:qso-all-fit}
}
    \tablewidth{0pt}
    \tablehead{\colhead{Source} & \colhead{$z$} & \colhead{$m_{\rm 1450\AA}$} & \colhead{$\chi_\nu^2$} & \colhead{$L_{\rm IR}$($10^{13}L_\odot$)} & \colhead{$f_{\rm host, IR}$} & \colhead{${\rm c_{SFR}}$} & \colhead{SFR($M_\odot/{\rm yr}$)} \\
    \colhead{(1)} &  \colhead{(2)} &
  \colhead{(3)} &
  \colhead{(4)} &
  \colhead{(5)} &
  \colhead{(6)} &
  \colhead{(7)} &
  \colhead{(8)} 
}
\startdata
SDSS J000239.39+255034.8  &	  5.80 &	19.0  &   1.32 & $\leq$3.18 & $\leq$0.44 &    0.88 & $\leq$2394 \\
SDSS J000552.34$-$000655.8*  &	  5.85 &	20.8  &   2.66 & $\leq$1.34 & $\leq$0.89 &    0.61 & $\leq$1408 \\
SDSS J001714.67$-$100055.4  &	  5.01 &	19.4  &   8.29 & $\leq$2.39 & $\leq$0.00 &    1.00 & $\leq$0.20(?) \\
SDSS J005421.42$-$010921.6  &	  5.09 &	20.5  &   1.31 & $\leq$2.32 & $\leq$0.68 &    1.00 & $\leq$3006 \\
SDSS J013326.84+010637.7*  &	  5.30 &	20.7  &   2.71 & $\leq$2.49 & $\leq$0.72 &    1.00 & $\leq$3425 \\
SDSS J020332.35+001228.6  &	  5.72 &	20.9  &   0.27 &      2.44 &    0.43 &    1.00 &    2028 \\
SDSS J023137.65$-$072854.5*  &	  5.41 &	19.5  &   8.64 & $\leq$2.66 & $\leq$0.54 &    1.00 & $\leq$2750 \\
SDSS J030331.40$-$001912.9*  &	  6.08 &	21.3  &   1.33 & $\leq$1.51 & $\leq$0.82 &    0.62 & $\leq$1467 \\
SDSS J033829.31+002156.3  &	  5.00 &	20.0  &   6.29 &      3.93 &    0.57 &    1.00 &    4305 \\
SDSS J035349.72+010404.4  &	  6.07 &	20.2  &   0.08 & $\leq$1.97 & $\leq$0.49 &    0.69 & $\leq$1281 \\
SDSS J073103.12+445949.4  &	  5.01 &	19.1  &   3.64 & $\leq$2.63 & $\leq$0.13 &    1.00 & $\leq$643 \\
SDSS J075618.14+410408.6  &	  5.09 &	20.1  &   1.41 &      3.73 &    0.69 &    1.00 &    4964 \\
SDSS J081827.40+172251.8  &	  6.00 &	19.3  &   1.60 &      3.13 &    0.16 &    1.00 &     971 \\
SDSS J083317.66+272629.0  &	  5.02 &	20.3  &   0.04 & $\leq$2.20 & $\leq$0.69 &    1.00 & $\leq$2901 \\
SDSS J083643.85+005453.3*  &	  5.81 &	18.8  &  29.14 & $\leq$4.19 & $\leq$0.19 &    1.00 & $\leq$1544 \\
SDSS J084035.09+562419.9  &	  5.84 &	20.0  &   0.42 &      2.72 &    0.67 &    1.00 &    3504 \\
SDSS J084119.52+290504.4  &	  5.96 &	19.6  &   0.41 & $\leq$2.10 & $\leq$0.42 &    0.69 & $\leq$1167 \\
SDSS J084229.23+121848.2  &	  6.06 &	19.9  &   6.80 & $\leq$3.66 & $\leq$0.34 &    0.61 & $\leq$1448 \\
SDSS J084627.85+080051.8  &	  5.04 &	19.6  &   0.59 & $\leq$2.44 & $\leq$0.63 &    1.00 & $\leq$2938 \\
BWE 910901+6942	          &       5.47 &	19.8  &   0.00 & $\leq$2.58 & $\leq$0.60 &    1.00 & $\leq$2979 \\
SDSS J090245.77+085115.8  &	  5.22 &	20.6  &   1.82 & $\leq$2.20 & $\leq$0.66 &    1.00 & $\leq$2785 \\
SDSS J091316.56+591921.5  &	  5.11 &	21.5  &   0.26 & $\leq$2.12 & $\leq$0.76 &    0.97 & $\leq$2996 \\
SDSS J091543.64+492416.7  &	  5.20 &	19.3  &   0.46 & $\leq$2.43 & $\leq$0.55 &    1.00 & $\leq$2557 \\
SDSS J092216.82+265359.1  &	  5.06 &	20.4  &   0.89 & $\leq$2.27 & $\leq$0.71 &    1.00 & $\leq$3099 \\
SDSS J092721.82+200123.7  &	  5.77 &	19.9  &   1.92 &      3.51 &    0.68 &    1.00 &    4561 \\
SDSS J095707.67+061059.5  &	  5.19 &	19.0  &   4.03 & $\leq$3.35 & $\leq$0.33 &    1.00 & $\leq$2151 \\
SDSS J101336.33+424026.5  &	  5.06 &	19.4  &   1.09 & $\leq$1.95 & $\leq$0.71 &    1.00 & $\leq$2647 \\
SDSS J103027.10+052455.0  &	  6.31 &	19.7  &   0.07 & $\leq$3.35 & $\leq$0.63 &    0.78 & $\leq$3165 \\
SDSS J104433.04$-$012502.2  &	  5.78 &	19.2  &   4.60 &      3.64 &    0.22 &    1.00 &    1530 \\
SDSS J104845.05+463718.3  &	  6.23 &	19.2  &   2.59 &      3.23 &    0.38 &    1.00 &    2360 \\
SDSS J111920.64+345248.2  &	  5.02 &	20.2  &   2.03 & $\leq$2.26 & $\leq$0.53 &    1.00 & $\leq$2278 \\
SDSS J113246.50+120901.7  &	  5.17 &	19.4  &   9.68 & $\leq$3.10 & $\leq$0.21 &    1.00 & $\leq$1262 \\
SDSS J113717.73+354956.9  &	  6.01 &	19.6  &   0.26 & $\leq$3.18 & $\leq$0.58 &    0.90 & $\leq$3207 \\
SDSS J114657.79+403708.7*  &	  5.01 &	19.7  &  20.25 & $\leq$2.16 & $\leq$0.26 &    1.00 & $\leq$1072 \\
SDSS J114816.64+525150.3  &	  6.43 &	19.0  &   4.23 &      5.48 &    0.36 &    1.00 &    3801 \\
RD J1148+5253	         &        5.70 &	23.1  &   0.00 & $\leq$2.58 & $\leq$0.93 &    0.96 & $\leq$4406  \\
SDSS J115424.74+134145.8  &	  5.08 &	20.9  &   1.76 & $\leq$2.29 & $\leq$0.59 &    1.00 & $\leq$2571 \\
SDSS J120207.78+323538.8  &	  5.31 &	18.6  &   2.93 &      4.79 &    0.44 &    1.00 &    4046 \\
SDSS J120441.73$-$002149.6  &	  5.03 &	19.1  &   4.33 &      4.42 &    0.54 &    1.00 &    4547 \\
SDSSp J120823.82+001027.7* &      5.27 &	20.5  &   0.29 & $\leq$1.99 & $\leq$0.91 &    0.90 & $\leq$3125 \\
SDSS J122146.42+444528.0  &	  5.19 &	20.4  &   6.62 & $\leq$2.35 & $\leq$0.37 &    1.00 & $\leq$1683 \\
SDSS J124247.91+521306.8*  &	  5.05 &	20.6  &   3.97 & $\leq$1.82 & $\leq$0.67 &    1.00 & $\leq$2344 \\
SDSS J125051.93+313021.9  &	  6.13 &	19.6  &  21.44 & $\leq$4.08 & $\leq$0.25 &    1.00 & $\leq$1954 \\
SDSS J130608.26+035626.3  &	  6.02 &	19.6  &   0.04 & $\leq$2.76 & $\leq$0.70 &    0.81 & $\leq$2992 \\
SDSS J133412.56+122020.7  &	  5.14 &	19.5  &   1.05 & $\leq$2.65 & $\leq$0.37 &    1.00 & $\leq$1869 \\
SDSS J133550.81+353315.8  &	  5.90 &	19.9  &   2.12 &      2.41 &    0.60 &    1.00 &    2762 \\
SDSS J133728.81+415539.9  &	  5.03 &	19.7  &   0.33 & $\leq$2.09 & $\leq$0.56 &    1.00 & $\leq$2261 \\
SDSS J134015.04+392630.8  &	  5.07 &	19.6  &   5.46 & $\leq$2.89 & $\leq$0.37 &    1.00 & $\leq$2031 \\
SDSS J134040.24+281328.2  &	  5.34 &	19.9  &   0.09 &      5.03 &    0.46 &    1.00 &    4404 \\
SDSS J134141.46+461110.3  &	  5.01 &	21.3  &   0.01 & $\leq$2.08 & $\leq$0.64 &    1.00 & $\leq$2558 \\
SDSS J141111.29+121737.4*  &	  5.93 &	20.0  &   7.51 & $\leq$2.21 & $\leq$0.64 &    0.68 & $\leq$1827 \\
SDSS J142325.92+130300.7  &	  5.08 &	19.6  &   0.12 & $\leq$2.47 & $\leq$0.37 &    1.00 & $\leq$1740 \\
FIRST J142738.5+331241    & 	  6.12 &	20.3  &   2.20 & $\leq$2.22 & $\leq$0.69 &    0.65 & $\leq$1910 \\
SDSS J143611.74+500706.9  &	  5.83 &	20.2  &   0.01 & $\leq$2.83 & $\leq$0.72 &    0.86 & $\leq$3346 \\
SDSS J144350.67+362315.2  &	  5.29 &	20.3  &  40.79 & $\leq$4.85 & $\leq$0.00 &    1.00 & $\leq$1.6(?) \\
SDSS J151035.29+514841.0  &	  5.11 &	20.1  &   0.14 & $\leq$2.34 & $\leq$0.44 &    1.00 & $\leq$1976 \\
SDSS J152404.10+081639.3  &	  5.08 &	20.6  &   0.02 & $\leq$2.22 & $\leq$0.75 &    1.00 & $\leq$3196 \\
SDSS J160254.18+422822.9  &	  6.07 &	19.9  &   2.62 & $\leq$3.25 & $\leq$0.32 &    0.68 & $\leq$1341 \\
SDSS J161425.13+464028.9  &	  5.31 &	20.3  &   7.25 & $\leq$3.20 & $\leq$0.28 &    1.00 & $\leq$1729 \\
SDSS J162331.81+311200.5  &	  6.25 &	20.1  &   1.46 & $\leq$2.59 & $\leq$0.43 &    0.99 & $\leq$2107 \\
SDSS J162626.50+275132.4  &	  5.30 &	18.7  &   7.64 &      5.35 &    0.03 &    1.00 &     292 \\
SDSS J162629.19+285857.6  &	  5.02 &	19.9  &   0.95 & $\leq$1.32 & $\leq$0.20 &    1.00 & $\leq$503 \\
SDSS J163033.90+401209.6  &	  6.07 &	20.6  &   0.49 & $\leq$1.87 & $\leq$0.78 &    0.63 & $\leq$1752 \\
SDSS J165902.12+270935.1  &	  5.32 &	18.8  &  11.12 & $\leq$3.76 & $\leq$0.21 &    1.00 & $\leq$1547 \\
SDSS J205406.49$-$000514.8  &	  6.04 &	20.6  &   1.47 &      3.15 &    0.41 &    1.00 &    2501 \\
SDSS J211928.32+102906.6  &	  5.18 &	20.6  &   1.49 & $\leq$1.40 & $\leq$0.32 &    1.00 & $\leq$857 \\
SDSS J222845.14$-$075755.2*  &	  5.14 &	20.2  &   6.81 & $\leq$1.34 & $\leq$0.62 &    1.00 & $\leq$1593 \\
WFS J2245+0024*           &       5.17 &	21.8  &   2.38 & $\leq$2.15 & $\leq$0.89 &    1.00 & $\leq$3685 \\
SDSS J231546.57$-$002358.1  &     6.12 &	21.3  &   1.41 & $\leq$1.94 & $\leq$0.76 &    0.61 & $\leq$1741 
\enddata
\tablecomments{
    Col. (1): Object name with a * is identified as a hot dust deficient (HDD) quasar;
    Col. (2): Redshift;
    Col. (3): Rest-frame 1450\AA~apparent magnitudes, or mag(1450\AA);  
    Col. (4): Reduced chi-square, as defined in \ref{sec:fitting-procedure};
    Col. (5): The integrated infrared luminosity (8-1000 $\mum$) from the fitted two-component SED models;
    if there are no any detections with $\lambda_{\rm rest} < 10\mum$, the
    derived AGN contribution is put as an upper limit; the host galaxy
    contribution is treated as an upper limit if there is no detection at
    $\lambda_{\rm rest} > 50\mum$;
    Col. (6): Relative contribution of the galaxy component to $L_{\rm IR}$.
    Col. (7): Correction of the host IR luminosity for the derivation of SFR;
    Col. (8): Star formation rate estimated from the galaxy component
    luminosity, assuming the Haro 11 star formation law; a question mark
    indicates the quasar has a minimal host contribution in the IR, in
    which case the derived SFR is dubious and not used in the analysis.\\
}
\end{deluxetable*}

We also searched for other $z\gtrsim5.0$ quasars with rest-frame far-IR
observations in the literature and found 33 more quasars not included in
\cite{Leipski2014} (hereafter {\it sample B}). The majority of them have been
listed in \cite{Calura2014}, except for RD J0301+0020 and TN J0924-2201. 
We collected all available observations on dust continuum as well as the
UV/optical continuum \citep{Bertoldi-Cox2002, Petric2003, Wang2008a, Wang2011a,
Wang2011b, Wang2013, Omont2013, Willott2013, Willott2015, Barnett2015}.
Because of the lack of constraints on the near-IR and mid-IR bands, we did not
make full fits as for the cases in {\it sample-A}, but scaled the templates to
some specific data points.  For the AGN component, considering the potential
extinction effect and possible lack of hot dust emission (to be discussed in
Section~\ref{sec:discussion}), we scale the template to the data point at
$\lambda_{\rm rest} = 0.1450 - 1.0\mum$, which yields a maximum AGN luminosity.
For the host component, we scale the Haro 11 template to the data point at
$\lambda_{\rm rest} > 50\mum$, which yields a minimum IR luminosity.  Then we
calculated their total infrared luminosities, the fraction of host
contribution, and star formation rates, as summarized in
Table~\ref{tab:qso-extended-fit}. Compared with {\it sample-A}, {\it
    sample-B} contains less-luminous quasars (mainly contributed by the
Canada-France High-z Quasar Survey, see \citealt{Omont2013} and references
therein), and consists almost entirely of quasars at $z>5.7$.

In the following discussion, we will mainly focus on {\it sample-A}, whose data
points are more uniformly collected and have the same detection limits.
We will discuss {\it sample-B} only as a complement to conclusions based on
{\it sample-A}.

\begin{deluxetable*}{lccccc}
    \tabletypesize{\scriptsize}
    \tablecaption{SED decomposition results for $z\gtrsim5.0$ 33 quasars ({\it sample-B})\label{tab:qso-extended-fit}
}
    \tablewidth{0pt}
    \tablehead{\colhead{Source } & \colhead{$z$} & \colhead{$m_{\rm 1450\AA}$}  & \colhead{$L_{\rm IR}$($10^{13}L_\odot$)} & \colhead{f$_{\rm host, IR}$} & \colhead{SFR($M_\odot/{\rm yr}$)} \\
    \colhead{(1)} & \colhead{(2)} &  \colhead{(3)} & \colhead{(4)} & \colhead{(5)} &  \colhead{(6)}
}
\startdata
SDSS J003311.40-012524.9 &   6.13 & 21.53 &       0.81 &        0.69 &  1079 \\
CFHQS J005006+344522 &   6.25 & 19.84 & $\leq$  2.99 & $\leq$0.46 & $\leq$2643 \\
CFHQS J005502+014618 &   6.02 & 21.82 &           0.28 &        0.33 &   179 \\
CFHQS J010250-021809 &   5.95 & 22.02 & $\leq$  1.27 & $\leq$0.88 & $\leq$2141 \\
SDSS J012958.51-003539.7 &   5.78 & 22.28 &       1.16 &        0.90 &  2013 \\
CFHQS J013603+022605 &   6.21 & 22.04 & $\leq$  3.09 & $\leq$0.65 & $\leq$3852 \\
CFHQS J021013-045620 &   6.44 & 22.25 &           0.19 &        0.25 &    92 \\
CFHQS J021627-045534 &   6.01 & 24.15 & $\leq$  1.11 & $\leq$0.98 & $\leq$2082 \\
CFHQS J022122-080251 &   6.16 & 21.98 & $\leq$  2.26 & $\leq$0.93 & $\leq$4016 \\
CFHQS J022743-060530 &   6.20 & 21.41 & $\leq$  1.08 & $\leq$0.73 & $\leq$1522 \\
SDSS J023930.24-004505.4 &   5.82 & 22.28 & $\leq$  1.66 & $\leq$0.93 & $\leq$2945 \\
RD J0301+0020 &   5.50 & 23.40 &          0.52 &        0.91 &   906 \\
CFHQS J031649-134032 &   5.99 & 21.72 & $\leq$  3.89 & $\leq$0.95 & $\leq$7077 \\
TN J0924-2201 &   5.20 & -- & $\leq$  1.04 & $\leq$0.84 & $\leq$1659.35 \\
CFHQS J105928-090620 &   5.92 & 20.75 & $\leq$  2.06 & $\leq$0.60 & $\leq$2377 \\
ULAS J1120+0641 &   7.08 & -- &           0.42 &        0.63 &   513 \\
ULAS J131911.29+095051.4 &   6.13 & 19.65 &       3.47 &        0.59 &  3927 \\
SDSS J142516.30+325409.0 &   5.85 & 20.62 & $\leq$  1.89 & $\leq$0.50 & $\leq$1810 \\
CFHQS J142952+544717 &   6.21 & 20.59 &           2.68 &        0.63 &  3246 \\
SDSS J150941.78-174926.8 &   6.12 & 19.63 & $\leq$  2.52 & $\leq$0.43 & $\leq$2057 \\
SDSS J162100.70+515544.8 &   5.71 & 19.89 & $\leq$  4.21 & $\leq$0.18 & $\leq$1438 \\
SDSS J164121.64+375520.5 &   6.04 & 21.19 & $\leq$  1.17 & $\leq$0.71 & $\leq$1603 \\
SDSS J205321.77+004706.8 &   5.92 & 21.20 & $\leq$  1.34 & $\leq$0.76 & $\leq$1946 \\
CFHQS J210054-171522 &   6.09 & 21.37 & $\leq$  3.46 & $\leq$0.24 & $\leq$1604 \\
SDSS J214755.40+010755.0 &   5.81 & 21.65 & $\leq$  1.44 & $\leq$0.65 & $\leq$1782 \\
CFHQS J222901+145709 &   6.15 & 21.90 &           0.19 &        0.06 &    22 \\
CFHQS J224237+033421 &   5.88 & 22.09 & $\leq$  1.48 & $\leq$0.90 & $\leq$2561 \\
SDSS J230735.35+003149.4 &   5.87 & 21.73 & $\leq$  1.68 & $\leq$0.45 & $\leq$1460 \\
SDSS J231038.88+185519.7 &   6.00 & 19.30 &       6.07 &        0.69 &  7996 \\
CFHQS J231802-024634 &   6.05 & 21.55 & $\leq$  1.38 & $\leq$0.82 & $\leq$2178 \\
SDSS J232908.28-030158.8 &   6.43 & 21.53 & $\leq$  1.38 & $\leq$0.00 & $\leq$   0.0(?) \\
CFHQS J232914-040324 &   5.90 & 21.96 & $\leq$  1.14 & $\leq$0.86 & $\leq$1886 \\
SDSS J235651.58+002333.3 &   6.00 & 21.77 & $\leq$  1.06 & $\leq$0.81 & $\leq$1653
\enddata
\tablecomments{
   Col. (1): Quasar name;
   Col. (2): Redshift;
   Col. (3): Rest-frame 1450\AA~AB apparent magnitudes, or mag(1450\AA); 
   Col. (4): The integrated infrared luminosity (8-1000 $\mum$) from the fit two-component SED models;
   if there is no detection with $\lambda_{\rm rest} < 10\mum$, the derived AGN
   contribution is put as an upper limit; the host contribution is treated as
   an upper limit if the quasar is not detected at $\lambda_{\rm rest} >
   50\mum$;
   Col. (5): Relative contribution of the galaxy component to $L_{\rm IR}$;  
   Col. (6): Star formation rate estimated from the galaxy component
   luminosity, assuming the Haro 11 star formation law; a question mark
   indicates the quasar has a minimal host contribution in the IR, in which
   case the derived SFR is dubious and not used in the analysis.\\
}
\end{deluxetable*}

\section{Discussion}\label{sec:discussion}

\subsection{Heating Sources for the Infrared Energy Output}\label{sec:heating}

To study the host galaxies of high-$z$ quasars from infrared SEDs, the heating
sources of the infrared-emitting dust and the contribution from the host star
formation should be examined first. Previously, a number of papers made the
assumption that the heating process for the FIR-emitting warm dust is dominated
by host star formation \citep[e.g.,][]{Leipski2014}, or assumed some
conservative fraction of host star formation heating
\citep[e.g.,][]{Wang2011b}.  From a theoretical point of view, \cite{Li2008}
and \cite{Schneider2014} studied the heating of the observed SED of J1148+5251,
an archetypal high-luminosity high-redshift quasar.  They argued that the
heating of the dust in the host galaxy could be dominated by processes related
to the central engine, rather than the host star formation. This field is quite
controversial.

As shown in Figure~\ref{fig:qso_all1}, the AGN template from \cite{Xu2015a} is
not sufficient to reproduce the far-IR SED of many $z \gtrsim 5$ quasars with
rest-frame far-IR detections. To investigate the average infrared properties of
these quasars, we fit the three stacked SEDs in \cite{Leipski2014}, shown in
Figure~\ref{fig:stacked_sed}. For the FIR-detected SED (from objects detected
at least three \herschel bands), we can see a clear contribution in the far-IR
from the host galaxy. For the partly detected SED (from objects with
significant PACS 100 $\mum$ and/or 160$\mum$ flux), the AGN template alone is
enough to reproduce the SED.  For the objects without any \herschel detections,
there are some HDD quasars with SEDs deviating from normal AGNs, as indicated
by the low ratio of rest 24 $\mum$ to optical.  The stacked SEDs for the 14
partly \herschel detected and the 33 non-detected systems show no evidence for
significant far-IR output over that of typical quasar templates. It would be
difficult to understand why just 10 of this sample had strong heating of the
host galaxy ISM by the quasar. A plausible explanation is that these 10 {\it
Herschel}-detected systems have high levels of star formation, while for the
other quasars the star formation is weak.

\begin{deluxetable*}{lcccccc}
    \tablewidth{0pt}
    \tablecolumns{7}
    \tablecaption{SED decomposition results for the stacked quasar SEDs\label{tab:stacked-fit}
		}
    \tablehead{
	\colhead{Source} & \colhead{N} & \colhead{Redshift} & \colhead{ L$_{\rm IR}$/$10^{13}$L$_{\odot}$} & \colhead{f$_{\rm host, IR}$} & \colhead{$\chi_{\nu}^2$} & \colhead{SFR($M_\odot/{\rm yr}$)} \\
	\colhead{(1)} & \colhead{(2)} & \colhead{(3)} & \colhead{(4)} & \colhead{(5)} & \colhead{(6)} & \colhead{(7)} 
    }
\startdata
FIR-detected  	 &    10 &  5.34  & 4.08   &   0.47   &    0.63  &        3666       \\
partly-detected  &    14 &  5.31  & 2.04   &   0.00   &    0.20  & $\lesssim1.5$ (?) \\
non-detected     &    33 &  5.20  & 0.87   &   0.00   &    0.61  & $\sim 0$ (?)
\enddata
\tablecomments{
    Results of full IR fits. Upper-limit data points are included in the
    evaluation process.\\
    Col. (1): Type of stacked SED;
    Col. (2): The number of stacked quasars;
    Col. (3): Average redshift; 
    Col. (4): The total infrared luminosity (8-1000 $\mum$) estimated from the ``Haro 11 +
    AGN'' two-component SED fit; 
    Col. (5): The fraction of luminosity of host template contribution to the
    whole fit SED, based on result from the ``Haro 11 + AGN''
    decomposition; 
    Col. (6): Reduced chi-square from the ``Haro 11 + AGN''
    decomposition; 
    Col. (7): Estimation of the star formation rate.
}
\end{deluxetable*}

    At very high redshift, cosmic microwave background (CMB) is also a source
    for dust heating \citep{Da-Cunha2013}. However, since the dust temperatures
    in high-$z$ quasar host galaxies are typically $\sim$ 35-50K (e.g.,
    \citealt{Xu2015a, Leipski2014}), at least twice the CMB temperature
    for the relevant redshift range ($T_{\rm CMB}\sim18$K at
    $z\sim5.5$), a correction is not significant compared with the other
    uncertainties in our derivations. 

\subsection{Are AGN Host Galaxies at $z\gtrsim 5$ Low-metallicity?}\label{sec:host-low-metal}

For quasars at $z\sim1-6$, emission line ratios are found to trace
(super-)solar gas metallicities (up to $\sim10\,Z_\odot$) in broad line regions
(BLRs) without any strong indication of redshift evolution \citep{Nagao2006,
Nagao2012, Jiang2007, Juarez2009}.  However, the mass of the BLRs is small
($10^2-10^4\,M_\odot$), and might not be representative of the overall
formation history of the galaxy. \cite{Wang-JM2010,Wang-JM2011} showed that the
star formation can be enhanced in the accretion flow of the AGN, possibly
resulting in locally increased metallicity. The narrow-line regions (NLRs) of
quasars at $z\sim1-4$ are also found to be around solar-metallicity without
strong evolution \citep{Matsuoka2009}. In contrast with the BLRs, the typical
size of the NLRs ($\sim10^{1-4}$ pc, \citealt{Bennert2006a, Bennert2006b}) is
comparable to the size of the host galaxies. The only quasar beyond $z \gtrsim
5$ with a NLR metallicity constraint is TN J0924$-$2201, a type-2 radio galaxy
at $z=5.19$ \citep{Matsuoka2011}. Considering the small sample size and
uncertainty of the metallicity calibration, the result for TN J0924$-$2201 does
not provide much knowledge of the metallicity in the $z \gtrsim 5$ quasars. We
do not have observational constraints on the metallicity of these quasar hosts
from emission line analysis.

Another possible approach to get metallicity constraints on (or near) distant
quasar hosts is from analyzing the absorbers with high H I content ($N_{\rm
H~I}\gtrsim10^{20} {\rm cm}^{-2}$), or so-called damped Lyman-alpha (DLA)
systems, at the redshift of the quasar \citep{Hennawi2009, Zafar2011}.
\cite{Hennawi2009} reported the discovery of a bright Lyman-$\alpha$ blob
associated with the $z=3$ quasar SDSS J124020.91+145535.6 and gave a lower
limit to the gas metallicity $Z\gtrsim 1/10\, Z_\odot$. \cite{Zafar2011}
studied a physical quasar pair Q0151+048 ($z\sim1.9$) and suggested an overall
metallicity of $0.01\,Z_\odot$ for a DLA associated with one member. The
redshifts of these two quasars are relatively low. It is also not clear if they
are representative of the general population. As argued by \cite{Finley2013},
statistical study shows the absorption of the associated DLAs is more likely to
happen in the galaxies neighboring the quasar, rather than in the AGN host
galaxy. Further detailed studies on larger samples are needed to make any
conclusive argument.

Several works argued that some massive galaxies at high-$z$ have solar
metallicity \citep[e.g.,][]{Maiolino2008, Mannucci2009,Rawle2014}. It is
possible that these objects are mature and highly evolved. However, we should
be cautious about the derived metallicity with very limited data points for
individual sources. Convincing measurements of metallicity at high-$z$ require
more understanding of the ISM in these systems. For $2.0 < z < 2.5$ galaxies,
statistical studies based on multiple metallicity tracers show that their
metallicities drop at large masses \citep{Maier2014,Cullen2014}.  For the very
early Universe, simulations suggest population III stars contribute little to
the chemical enrichment of the ISM \citep{Valiante2009}.  The existence of a
huge population of low-metallicity systems between the cosmic reionization and
$z > 2.5$ should be expected.  Since high-$z$ quasars are originally identified
by their AGN features, the properties of their host galaxies should not be much
biased by the selection.  It is therefore plausible that the high-$z$ quasar host
galaxies have metallicities moderately, if not substantially, below solar.

In this work, hints for the low-metallicity of the AGN host galaxies at
$z\gtrsim5$ are from the successful reproduction of the observed SEDs based on
two-component fits, as shown in Section~\ref{sec:demonstration}. The high dust
temperature and boosted mid-IR emission are two major features of the IR SED of
Haro 11, a dwarf galaxy with metallicity $Z\sim1/3\,Z_\odot$. For the IR SED of
these quasars, the low-metallicity Haro 11 template works significantly better
than the normal SF templates.

\cite{Xu2015a} discovered a warm mid-IR component of some type-1 quasars at
$z\sim0.7-2.5$, which can not be reproduced by the combination of the AGN
template and normal SF template. This warm excess is found to be more prominent
at higher redshifts in their sample. As shown in
Section~\ref{sec:demonstration}, a strong mid-IR SED excess also does exist
when fitting the host galaxy with normal SF templates for the $z\gtrsim5$
quasars. In contrast, by introducing the Haro 11 template, the mid-IR part of
the SED of the $z\gtrsim 5.0$ quasars is reproduced better: there is no strong
hint of the warm excess for the majority of the quasars. The low-metallicity of
the host galaxy is a possible explanation for many such warm excesses at
high-$z$: the dust population in the low-metallicity environment tends to be
dominated by small-size grains, which would result in substantial emission in
the mid-infrared. Due to the increase of the mid-IR emission, the effective
dust temperature fit from the whole infrared SED is also boosted.  Nonetheless,
a small number of $z\gtrsim5$ quasars still show a mid-IR warm excess, such as
J0338+0021 and J1602+4228, even with the Haro 11 template fitting.  We
suggest that such additional warm excess not reproduced by the ``Haro 11 +
AGN'' SED model could be due to an extreme circumnuclear starburst or that the
host galaxy has a much lower metallicity.

\subsection{The Star Formation Rates of Quasars at $z \gtrsim 5$}\label{sec:qso-sfr}

In estimating a SFR, the largest uncertainty comes from the assumed star
formation calibration. In Appendix~\ref{sec:dwarf-sfr-calibration}, the star
formation determination for the low-metallicity dwarf galaxies is discussed.
We show that the \cite{Kennicutt1998} IR star formation law is still
valid to roughly estimate the obscured star formation rates for the
low-metallicity dwarf galaxies, including Haro 11. Besides the obscured star
formation, we also consider the unobscured star formation as revealed by the UV
emission. As shown in Appendix~\ref{sec:dwarf-sf-haro11}, Haro 11 has a low UV
star formation rate estimate, which is only $\sim$10\% of that deduced from the
far-IR. For a $2000\,M_\odot/{\rm yr}$ infrared SFR, the corresponding UV SFR
would be $200\,M_\odot/{\rm yr}$, consistent with the upper limit given for the
archetypal $z\sim6$ quasar J1148+5251 \citep{Mechtley2012}. The UV star
formation of high-luminosity quasar host galaxies at $z\sim2.6$ is also found
to be quite weak \citep{Cai2014}. These examples indicate that a low
contribution to the estimated SFR from the UV is appropriate for quasar
host galaxies identical to those for J1148+5251 and the \cite{Cai2014}
quasar sample. However, we can not rule out the possibility that some of
the host galaxies at the epoch of reionization have larger escape fractions
than Haro 11 and hence a large fraction of UV emission, causing us to
underestimate their total SFRs.

Many censored data points also make the SFR estimation difficult. For the
quasars with at least two detections in the far-IR, the host galaxies are
reasonably well fit. The derived star formation rates are on the order of
$10^3\,M_\odot/{\rm yr}$, a typical value also found by other authors
\citep[e.g.,][]{Wang2008a, Leipski2014}. For quasars without any far-IR
detections, we could only determine the upper-limits of their SFRs.  As
described in Section~\ref{sec:fitting-procedure}, we consider all the censored
data points during the fitting process. For sources without far-IR
detections, the fitted upper limits on the SFRs are based on templates
constrained by multiple 3-$\sigma$ non-detections and result in overestimated
SFR constraints. To solve this problem, we scale the host template to each non-detected
3$\sigma$-limit observation, derive the respective SFR, and pick the lowest
one as the final constraint on the quasar host SFR. During this process, the
contribution of the fitted AGN component is fixed and subtracted when deriving
the SFR.

For the HDD quasars, the AGN template fails at $\lambda \gtrsim 1.0\mum$. In
addition, none of them are detected in the far-IR. To derive conservative upper
limits for their SFRs, we assume that all their far-IR emission comes from the
host galaxy. Consequently, we ignore the fitted AGN component when we scale the
host template to the observations.

We ignore the results for J0017$-$1000 and J1443+3623 of {\it sample-A}, whose
host contribution is too minimal to be evaluated. From Kaplan-Meier
analysis\footnote{As implemented in the Astronomy Survival Analysis Package
(ASURV, \citealt{Lavalley1992})}, the mean infrared host galaxy luminosity
of the {\it sample-A} is
\begin{equation}
    \langle \log (L_{\rm SF,IR}/L_\odot) \rangle  = 12.51\pm0.10   \tag*{(sample A)} ~,
\end{equation}
which corresponds to an average star formation rate
\begin{equation}
    \langle {\rm SFR}\rangle  = 621^{+161}_{-128} \,M_\odot/{\rm yr}  \tag*{(sample A)} ~.
\end{equation}

Another approach to compute the average star formation rate is to analyze the
stacked SEDs.  As shown in Figure~\ref{fig:stacked_sed} and
Table~\ref{tab:stacked-fit}, the fraction of the host contribution is too small
to give any physical constraints on the SFR of the stacked SEDs of {\it
Herschel} partly- and non-detected quasars. We simply conclude substantial star
formation only happens in the \herschel FIR-detected stacked SED, whereas the
star formation in other stacked SEDs is minimal and set to be zero. Then an
arithmetic mean of {\it sample-A} is 
\begin{equation}
    \langle {\rm SFR}\rangle  = 643\, M_\odot/{\rm yr}  \tag*{(A-stacked)} ~.
\end{equation}
This result is almost the same as that from the Kaplan-Meier analysis for
individual sources, confirming the validity of the result from the Kaplan-Meier
estimator.

For {\it sample-B}, after rejecting SDSS J2329-0301 due to its minimal
host contribution, we use the Kaplan-Meier approach to derive a mean infrared
luminosity for 32 quasars 
\begin{equation}
    \langle \log (L_{\rm SF,IR}/L_\odot) \rangle  = 12.27\pm0.22   \tag*{(sample B)} ~,
\end{equation}
which corresponds to an average star formation rate
\begin{equation}
    \langle {\rm SFR}\rangle  = 357^{+236}_{-142} \,M_\odot/{\rm yr}  \tag*{(sample B)} ~.
\end{equation}
This estimate is subject to systematic errors because the majority of the {\it sample-B}
members only have submillimeter measurements at 1.2 mm, and these fall well beyond
the peaks of their far-IR SEDs. Therefore, any deviation of the SED from the
template will result in significant errors in the estimate of infrared
luminosity. Nonetheless, within the errors, this $\langle {\rm SFR}\rangle$ is
similar to that from {\it sample-A}. In fact, we will show in Section
\ref{sec:bh-galaxy-evolution} that the indicated slightly lower SFRs for {\it
sample-B} is as might be expected from the generally lower luminosities of
their AGNs.

We believe that the average SFR estimated above is robust even if Haro 11 is not
representative for some quasar host galaxies. As shown in
Section~\ref{sec:test-galaxy}, the results from the Haro 11 template are not
substantially different from the normal SF templates in \cite{Rieke2009}. 
    In fact, the AGN template is principally fixed by data points at
    $\sim1-5~\mum$, leaving the SF template to be matched to the mid-IR to
    far-IR SED. The large range between the maximum star formation rates and
    the averages suggests that star formation is very ``bursty" in the host
    galaxies, and that the averages can be considered to represent the rates
    integrated over time. These issues are discussed in
    Section~\ref{sec:bh-galaxy-evolution}.

\subsection{AGN Luminosity}

\begin{figure}[htbp] 
    \begin{center} 
	\includegraphics[width=0.9\hsize]{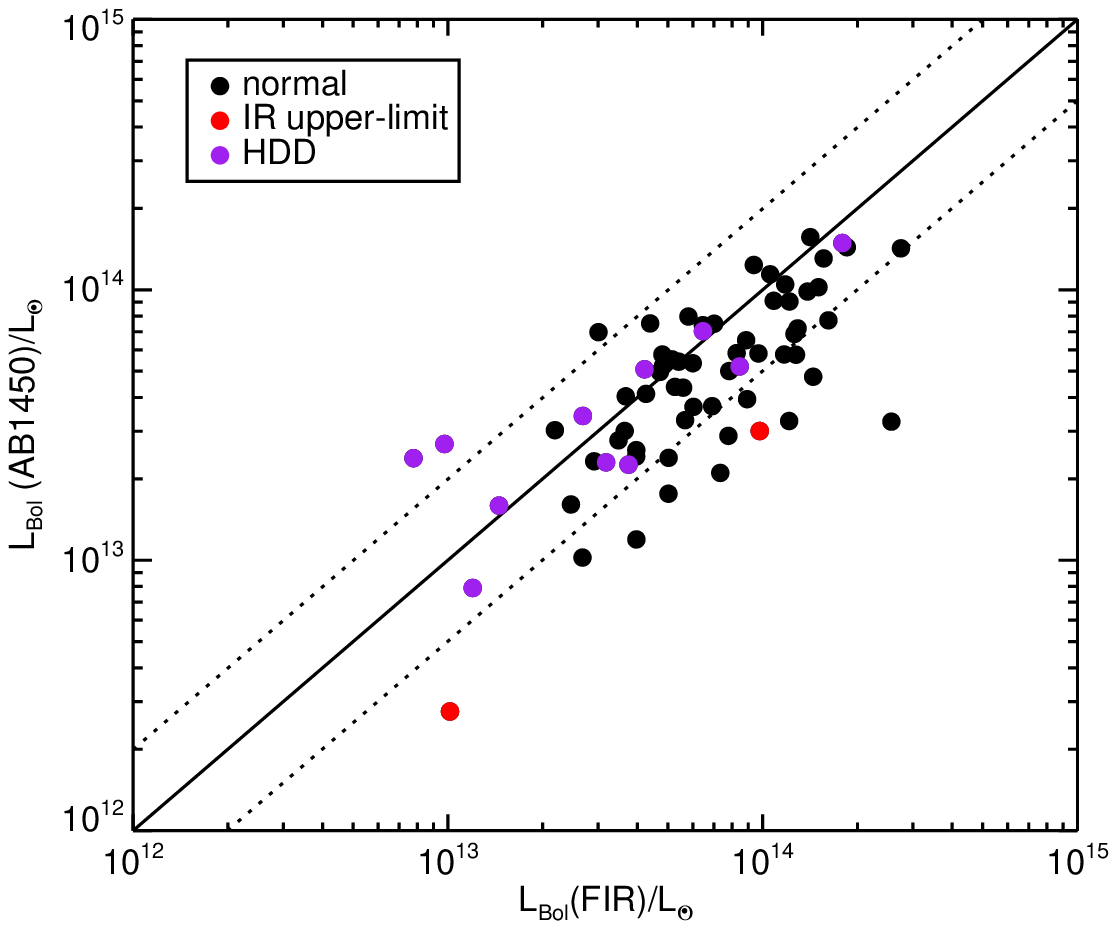} 
	\caption{
	    Comparison of AGN luminosities derived from mag(1450\AA) and infrared
	    SED fits. 0.3 dex offsets are shown as dotted lines.  Purple
	    dots are for the 11 HDD quasars, red dots for two quasars
	    with upper-limit of $L_{\rm AGN,IR}$, black dots for other quasars.
	} 
	\label{fig:AGN_lum_comp} 
    \end{center} 
\end{figure}

The total AGN luminosity can be estimated from integrating the Elvis template
(e.g., \citealt{Hao2014}, \citealt{Xu2015a}.  Since our fits are limited to the
infrared (1-1000 $\mum$), the total AGN luminosity can be derived by scaling an
infrared-to-bolometric correction of 5.28 \citep{Xu2015a} to the AGN total
infrared luminosity $L_{\rm IR,AGN}$\footnote{For quasars with $L_{\rm IR}$
upper limits in Table~\ref{tab:qso-all-fit}, we can still get their AGN total
infrared luminosities: $L_{\rm IR, AGN}= L_{\rm IR}(1-f_{\rm host,IR})$, where
$f_{\rm host,IR}$ is the host galaxy contribution upper limit.}. Before
that, we check the validity of the Elvis template in the UV/optical bands. As
shown in Figure~\ref{fig:stacked_sed}, the \cite{Elvis1994} template reproduces
the UV to mid-IR stacked SED of these quasars well. For individual sources,
although there are some detailed offsets, the Elvis template generally matches
the observations. The monochromatic flux at rest frame 1450\AA~is a frequently
used indicator of AGN UV continuum brightness in the literature. By applying a
scaling factor of 4.65\footnote{This value is derived based on the
\cite{Richards2006} template.} on the $\nu L_{\nu}$(1450\AA), the AGN
bolometric luminosity can be estimated.  Taking mag(1450\AA) in the
literature as a crude but independent tracer of AGN bolometric luminosity,
in Figure~\ref{fig:AGN_lum_comp}, we plot the mag(1450\AA)-based AGN
bolometric luminosities against the $L_{FIR}$-based ones. There is a small
offset from 1:1 on the correlation between the AGN luminosities from
mag(1450\AA) and from the infrared bolometric correction, which can be
explained by possible UV extinction. In summary, though the normalization of
the AGN template is constrained by the rest-frame near-IR to mid-IR data
points, the residuals of observed UV/optical SEDs from the IR fit to the Elvis
template are generally small. 

\cite{Leipski2014} pointed out 11 quasars with a dearth of very hot dust.  We
confirm their peculiarity by comparing their observed SEDs with the Elvis
template.  If normalized at UV/optical wavelengths, the Elvis template clearly
overestimates the observed SED beyond rest frame 1 $\mum$. Since we do not have
a clear picture of the full wavelength SED of these HDD objects, their
luminosities are hard to derive. We still rely on the $L_{\rm FIR}$-based
luminosity, rather than UV-based luminosity, for two reasons: (1) the
UV-optical SED could suffer extinction, thus underestimating the total
bolometric luminosity; (2) the UV emission is not isotropic while the sources
are optically thin in the near- and mid-IR \citep{Marconi2004}.  Judging by the
UV/optical observation, we do not expect the template-derived luminosity has
more than one order of magnitude deviation from the observed one. In
Figure~\ref{fig:AGN_lum_comp}, it is interesting to note the HDD quasars in
this work generally follow the same trend as normal quasars.

Finally, we can derive the average AGN bolometric luminosity of {\it sample-A}
with standard error to be:
\begin{equation}
    {\langle L_{\rm AGN}\rangle  = 5.287\times\langle L_{\rm AGN,IR}\rangle  } \approx (7.48\pm 0.62)\times 10^{13} \, L_\odot ~. \tag*{(sample A)}
\end{equation}

Similarly, {\it sample-B} has
\begin{equation}
    {\langle L_{\rm AGN}\rangle  \approx (2.63\pm 0.14)\times 10^{13}\, L_\odot}    ~.  \tag*{(sample B)}
\end{equation}
This value is $\sim36\%$ of {\it sample-A}

\subsection{Implications for BH-galaxy Evolution}
\label{sec:bh-galaxy-evolution}

\begin{figure}[htp] 
    \begin{center} 
	\includegraphics[width=0.9\hsize]{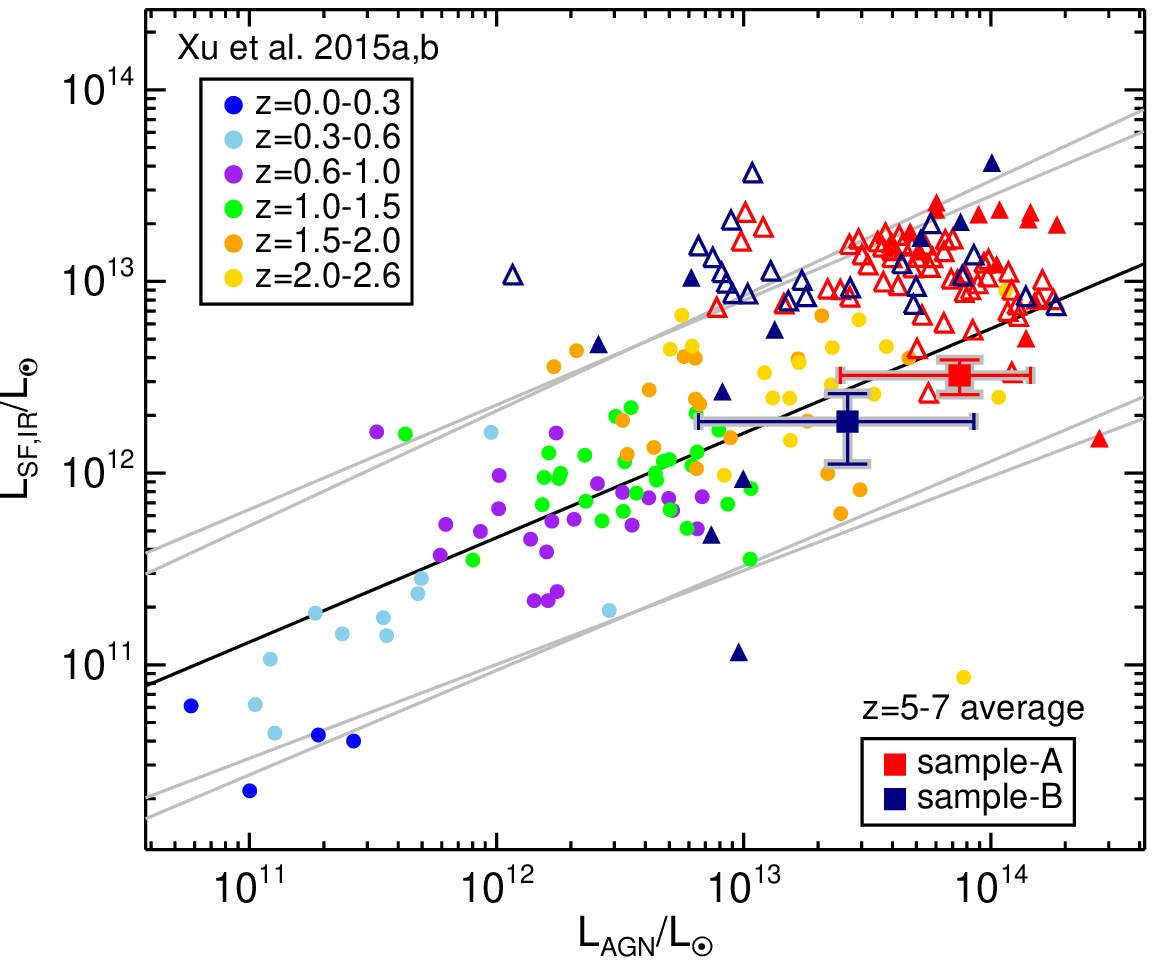} 
	\caption{
	    The IR luminosity of the star-formation component $L_{\rm SF,IR}$
	    versus the total AGN luminosity $L_{\rm AGN}$ for quasars at
	    different redshifts. Only type-1 AGNs in \cite{Xu2015a} are plotted
	    (small dots). The solid back line is the best-fit unweighted
	    relationship for all type-1 AGN in \cite{Xu2015a} with a slope of
	    0.55, with shifted relation by 1-$\sigma$ error of either slope or
	    intercept shown as grey solid lines. Our measurements for
	    individual $z\gtrsim5$ quasars are also shown as red ({\it
	    sample-A}) and blue ({\it sample-B}) triangles, with filled for
	    detections and open for upper-limits. We also plot the 
	    Kaplan-Meier mean of $z\gtrsim5$ AGN host galaxy IR luminosities at
	    mean AGN luminosities for {\it sample-A} (red square) and {\it
	    sample-B} (blue square), with horizontal error bars showing the
	range including 80\% of the sample's sources.
	} 
	\label{fig:QSO_AGN_SF} 
    \end{center} 
\end{figure}

We now compare the relative strength between SF activities and AGN luminosities
of $z\gtrsim 5$ quasars with that of relatively low-$z$ and intermediate-$z$
quasars. In Figure~\ref{fig:QSO_AGN_SF}, we put the average values for
$z\gtrsim 5$ quasars on the relation between SF IR luminosity, $L_{\rm SF,IR}$,
and AGN luminosity, $L_{\rm AGN}$, for the \cite{Xu2015a} type-1 quasar sample.
The average properties of the $z\gtrsim 5$ quasars fall along the fit relation.
This is unexpected, however, since quasars at $z\gtrsim 5$ and those at $z < 3$
should be in different star formation phases.  From a theoretical perspective,
there should be no star formation main sequence as is the case in the $z < 3$
Universe, but bursts of star formation and periods of near-zero star formation
rates, likely due to the dynamically disturbed gas within the galaxy halo
\citep[e.g.,][]{Muratov2015}. Current observations suggest the star formation
in some $z > 5$ quasars is extremely vigorous with SFRs at levels of $\gtrsim
10^3~M_{\odot}/{\rm yr}$ \citep[e.g.,][]{Wang2008a} or relatively mild with
SFRs $\lesssim$~50$~M_{\odot}/{\rm yr}$ \citep[e.g.,][]{Willott2013}.  Despite
this large dispersion, an underlying relation between the average host star
formation and AGN luminosity, which has been suggested for very luminous AGNs
at $z\lesssim2.5$ \citep[e.g.,][]{Netzer2009, Rosario2012}, seems to already
exist at $z\sim5-6$.

By taking an average of the star formation rates of quasar host galaxies in the
largest sample at $z\sim5-7$, we can retrieve a rough time-averaged star
formation rate during the lifetime of these quasar host galaxies.  In other
words, we assume that the relative number of host galaxies with very active
star formation (SFR$\sim10^3~M_\odot/{\rm yr}$) to those without significant
star formation (SFR$\lesssim10~M_\odot/{\rm yr}$) reflects the relative time
duration of the star-bursting phase to the quiescent phase of the galaxies.
Since stars lose mass quickly after leaving the main sequence, we can ignore
their contribution to the stellar mass of the galaxy. The main sequence
lifetime of a star with mass $M$ is
\begin{equation}
    \tau _{\rm MS} = 10\times(M/M_\odot)^{-2.5} ~~{\rm Gyr} ~.
\end{equation}
Then we have a simple model relating
the stellar mass $M_*$ to the galaxy growth time
$\Delta t$ and the initial mass function $\xi (m)$ as:
\begin{equation}\label{eqn:mass_int}
    \langle M_{*}\rangle  \sim \int^{\Delta t}_{0}  \int^{M_{\odot}(t/10)^{-0.4}}_{M_{\rm low}} \xi (m) \langle {\rm SFR}\rangle  \mathrm{d}m \mathrm{d} t  ~,
\end{equation}
where $M_{\rm low}$ is the minimum stellar mass, which is assumed to be
$0.1\,M_\odot$; $M_{\odot}(t/10)^{-0.4}$ is the maximum stellar mass of stars
that are still on the main sequence at time $t$ (unit: Gyr); $\Delta t$ is the epoch
of galaxy mass assembly.

To start, we simply assume a standard Salpeter IMF \citep{Salpeter1955}, and
estimate the increase in host galaxy stellar mass since the start of the cosmic
reionization, i.e., $z=8.8^{+1.7}_{-1.4}$ \citep{Planck2015}, corresponding to
a Universe age $0.573^{-0.122}_{+0.150}$ Gyr. Then the time duration $\Delta t$
from reionization to the average age of the quasars in our sample ($\langle
z\rangle$=5.5) is $\Delta t = 0.489 ^{+0.122}_{-0.150}$ Gyr. Taking the average
star formation rate for {\it sample-A} and solving the integration in
Equation~\ref{eqn:mass_int}, we derive the stellar mass that could form in a
$z\sim5.5$ quasar to be
\begin{equation}
    \langle M_{*}\rangle  \approx 3.0^{+0.7}_{-0.9} \times 10^{11}\, M_\odot ~.
\end{equation}

This rough estimate is not highly sensitive to the starting redshift, nor to
the form of the standard IMF (see Table~\ref{tab:galaxy-mass}). For example, if
we assume a start at $z=20$, the total stellar mass increases by less than a
factor of two. We can also change the power-law Salpeter IMF to a more
realistic \cite{Kroupa2003} IMF with a turnover below $\sim 1~M_\odot$ and use
the updated IR star formation calibration in \cite{Kennicutt2012}. The derived
$M_*$ is lower by up to 10\%. As a result, we estimate that the SFRs we deduce
for the host galaxies are likely to result in formation of a net mass
$~(3-5)\times10^{11} M_\odot$ during the major assembly phases for these quasar
host galaxies. Although this result was derived for {\it sample-A}, the high
average SFR also exhibited for {\it sample-B} indicates it is generally true
for high-redshift quasars.

We compare this mass with typical masses for the central black holes in these
galaxies, $\sim 10^8-10^9~M_\odot$ (e.g., \citealt{Jiang2007, Kurk2007,
Kurk2009, Willott2010a, De-Rosa2011, Jun2015}). It appears that the formation
of stellar mass is adequate to establish $M_{\rm BH}/M_{*,total}$ at 0.1 - 1\%,
which is in good agreement with estimates of this parameter locally (e.g.,
\citealt{Kormendy-Ho2013} and references therein), and with the evidence at
$z\lesssim2$ that the ratio does not evolve substantially with redshift. If
some of the $z\gtrsim5$ host galaxies have a larger escape fraction than Haro
11 and hence their SFRs are under-represented by our approach, this conclusion
is strengthened. Our conclusion agrees with the measurement of the dynamical
masses of these systems \citep{Willott2015} that also indicates very little
evolution of this ratio with redshift up to z$\sim$6.

\begin{deluxetable}{lccc}
    \tablecaption{Estimation of a typical host galaxy mass\label{tab:galaxy-mass}}
\tablewidth{0pt}
\tablehead{
    \colhead{IMF} & \colhead{SFR ($M_\odot/{\rm yr}$)$^a$} & \colhead{ $M_*$ ($M_\odot$)$^b$} & \colhead{ $M_*$ ($M_\odot$)$^b$} \\
    \colhead{} & \colhead{} & \colhead{ $z_{0}=8.8$} & \colhead{ $z_{0}=20$} 
}
\startdata
Salpeter            & 621       & $3.02\times10^{11}$ & $5.42\times10^{11}$  \\
Kroupa              & 534$^c$   & $3.00\times10^{11}$ & $4.80\times10^{11}$  
\enddata
\tablecomments{
    $^a$ Values derived from {\it sample-A};
    $^b$ Stellar masses derived based on Equation~\ref{eqn:mass_int}, $z_{0}$ is the redshift when the first galaxies begin to assembly their masses;
    $^c$ \cite{Kennicutt2012} updated their original IR star formation
    calibration \citep{Kennicutt1998} with the \cite{Kroupa2003} IMF. The SFR should
    be reduced to $\sim$86\%.
}
\end{deluxetable}

\section{Summary}\label{sec:summary}

In this work, we describe an effective strategy to analyze the infrared output
of high-$z$ quasars. A two-component SED model to quantify and distinguish star
formation and nuclear activity is proposed: the host galaxy component can be
represented by the SED of Haro 11, a low-metallicity dwarf galaxy with extreme
compact star-forming regions; the AGN component can be represented by the
\cite{Elvis1994} AGN template with the IR star formation contribution
subtracted. Such a simple model can help us have a better idea of the AGN and
star formation contribution for these high-$z$ objects when only limited
observations are available. Using this method, we have analyzed the AGN
contribution and host galaxy contribution to the infrared SEDs of 69 quasars
with \herschel observations in \cite{Leipski2014} and another 33 quasars in the
literature. Our major conclusions are as follows:

~

1. Haro 11 is a faithful representation for the host galaxies of $z\gtrsim5$
quasars. Besides its moderately low-metallicity, Haro 11 has a very high
star formation surface density, which is common for high-redshift galaxies as
well as quasar hosts.

2. The AGN contribution to the UV-to-IR SEDs of $z\gtrsim5$ quasars can be
well-represented by the \cite{Elvis1994} AGN template with the star formation
contribution to the IR subtracted. For the infrared, except for the HDD
quasars, there is no detectable over-prediction of the observed flux from the
modified AGN template. After subtracting the AGN contribution in the IR, the
residual flux can be interpreted as the IR contribution from host galaxy star
formation, which is well-fit by the Haro 11 template.

3. The warm excess found for some high-$z$ quasars (see, e.g., \citealt{Xu2015a})
can be produced by the introduction of the Haro 11 template, suggesting this
feature may reflect the low-metallicity of the quasar host. The Haro 11
template also shares the high dust temperature found in the far-IR measurements
of high-$z$ quasars. That is, these two features can be explained by the dust
properties and distribution in the low-metallicity environment.

4. The average SFR of the $z\sim5-6.5$ quasars observed by {\it Herschel}
is $\sim620\,M_\odot/{\rm yr}$, considering both far-IR detected and
non-detected observations. 

5. For the overall population of $z\gtrsim5$ quasars, the  total AGN
luminosity $\langle L_{\rm AGN}\rangle$ and average infrared luminosity from
star formation $\langle L_{\rm SF,IR}\rangle$ fall along the relation defined
by $z<2.6$ quasars. This is unexpected since the star formation at $z\gtrsim5$
and that at $z\lesssim3$ should experience different phases (e.g., bursty vs.
relatively steady).

6. Assuming the sample averaged SFR is a rough time-averaged SFR estimate
during the lifetime of the quasar host galaxies, we found the $z\sim5-6$ quasar
host galaxies could form $\sim(3-5)\times10^{11}M_\odot$ of stars.  With the
black hole mass measurements of these quasars, such massive host galaxies make
possible a local BH-galaxy mass relation $M_{\rm BH}/M_*\sim0.1-1\%$ at
$z\sim6$, suggesting there may not be strong redshift evolution of the
BH-galaxy mass ratio.

\acknowledgments

We are grateful to the referee for comments and suggestions that improved the
clarity of this paper. We thank Eiichi Egami, Richard Green, Daniel Stark and
Xiaohui Fan for helpful suggestions. This work was supported by NASA grants
NNX13AD82G and 1255094.  This publication has made use of data products from
the {\it Wide-field Infrared Survey Explorer}, which is a joint project of the
University of California, Los Angeles, and the Jet Propulsion
Laboratory/California Institute of Technology, funded by the National
Aeronautics and Space Administration.  This research has also benefited from
the use of the NASA/IPAC Extragalactic Database (NED) which is operated by the
Jet Propulsion Laboratory, California Institute of Technology, under contract
with the National Aeronautics and Space Administration.

\bibliographystyle{apj.bst}

~


\appendix

\section{Templates for Dwarf Galaxies}\label{sec:low-metal-template}

\subsection{Sample and Infrared Data}

The sample used to compute the SEDs of low-metallicity galaxies is from the
Dwarf Galaxy Survey \citep[DGS;][]{Madden2013}. This sample covers the
full metallicity range observable in the local Universe with
\metalOH~ranging from 7.14 to 8.43, and spans four orders of magnitude in star
formation rates. 

The far-IR data adopted here is mainly from \cite{Remy-Ruyer2013}
\footnote{\cite{Remy-Ruyer2015} corrected the PACS photometry for three
galaxies, HS 0822+3542, HS 1442+4250 and Tol 0618. However, none of our
arguments made in the appendix would be changed.}. A total of 48 dwarf
galaxies were observed with PACS and SPIRE on board of the {\it Herschel Space
Observatory} at 70, 100, 160, 250, 350, and 500 $\mum$. For I~Zw~18, we update
with the \herschel data from \cite{Fisher2014}. We also collect the near-IR to
mid-IR photometry data for the whole sample from the {\it Wide-field Infrared
Survey Explorer} (WISE ,\citealt{wise}) at 3.4, 4.6, 12 and 22 $\mum$.  47 of
these galaxies were found to have low-resolution mid-IR spectroscopic
observations from the \spitzer archive. We collect the staring-mode \spitzerirs
spectra from the Cornell Atlas of \spitzerirs Sources
\citep[CASSIS;][]{Lebouteiller2011}\footnote{The Cornell Atlas of Spitzer/IRS
Sources (CASSIS) is a product of the Infrared Science Center at Cornell
University, supported by NASA and JPL.} and adopt the post-BCD products from
the \spitzer Heritage Archive (SHA) for mapping-mode observations, which were
reduced in the SSC Pipeline Version S18.18. The latter spectral maps are
combined into a single spectrum to represent the mid-IR emission continuum of
the galaxy.  However, since the surface brightnesses of dwarf galaxies are
typically low, only a few \spitzer spectra have high enough signal-to-noise
ratio (S/N) continuum to be useful for our derivation of the full IR SEDs.
Finally, we focus on 19 DGS galaxies to study their IR SEDs.

\subsection{SED Modeling and Template Construction\footnote{After the
submission of this paper, a comprehensive study of the physical basis of the
infrared SEDs of the DGS sample of galaxies was published by
\cite{Remy-Ruyer2015b}. Their conclusions about the general shape of the SEDs
are similar to ours.}}

In \cite{Remy-Ruyer2013}, a simple modified blackbody is fit to the far-IR
photometry of the DGS sample to derive dust properties like temperature, mass,
and emissivity index.  After introducing the WISE mid-IR data at 12 and 22
$\mum$, the far-IR SED peak shifts towards a shorter wavelength for the
majority of sources, leading to a higher dust temperature. One single
(modified) blackbody is not enough to represent the full IR SED, since galaxies
always have a range of dust temperatures \citep[e.g.,][]{Dunne2001,
Willmer2009, Galametz2012, Kirkpatrick2012}

Here we utilize the \cite{Casey2012} procedure (hereafter CMC fits) to address
the mid-IR excess.  \cite{Casey2012} developed a fitting routine to fit the IR
data points with a modified blackbody plus a power-law component. The mid-IR
component is described as an analytical function
\begin{equation}
    S(\lambda)_{\rm MIR} = N_{\rm MIR} \lambda^{\alpha} e^{-(\lambda/\lambda_{\rm c})^2} ~,
\end{equation}
where $\alpha$ is the mid-IR power-law slope, and $\lambda_{\rm c}$ the
power-law turnover wavelength. The normalizing factor $N_{\rm mid-IR}$ and turnover
wavelength $\lambda_{c}$ are bounded with other parameters. We relax the
bounding condition of $\lambda_{c}$ to fit the diverse SEDs of the DGS sample,
leaving other configurations unchanged \citep[see][]{Casey2012}.  Since the
light from old stellar populations may contribute to the SED (mainly at short
wavelengths), we assume the emission in the WISE W1 band is completely stellar
and scale a Rayleigh-Jeans tail to estimate the stellar contribution to other
bands. The final SED fit is done for $\lambda > 8\mum$ data points after
subtracting the possible old population stellar contribution from the observed
fluxes.

\begin{figure*}[htp] 
    \begin{center} 
	\includegraphics[width=0.9\hsize]{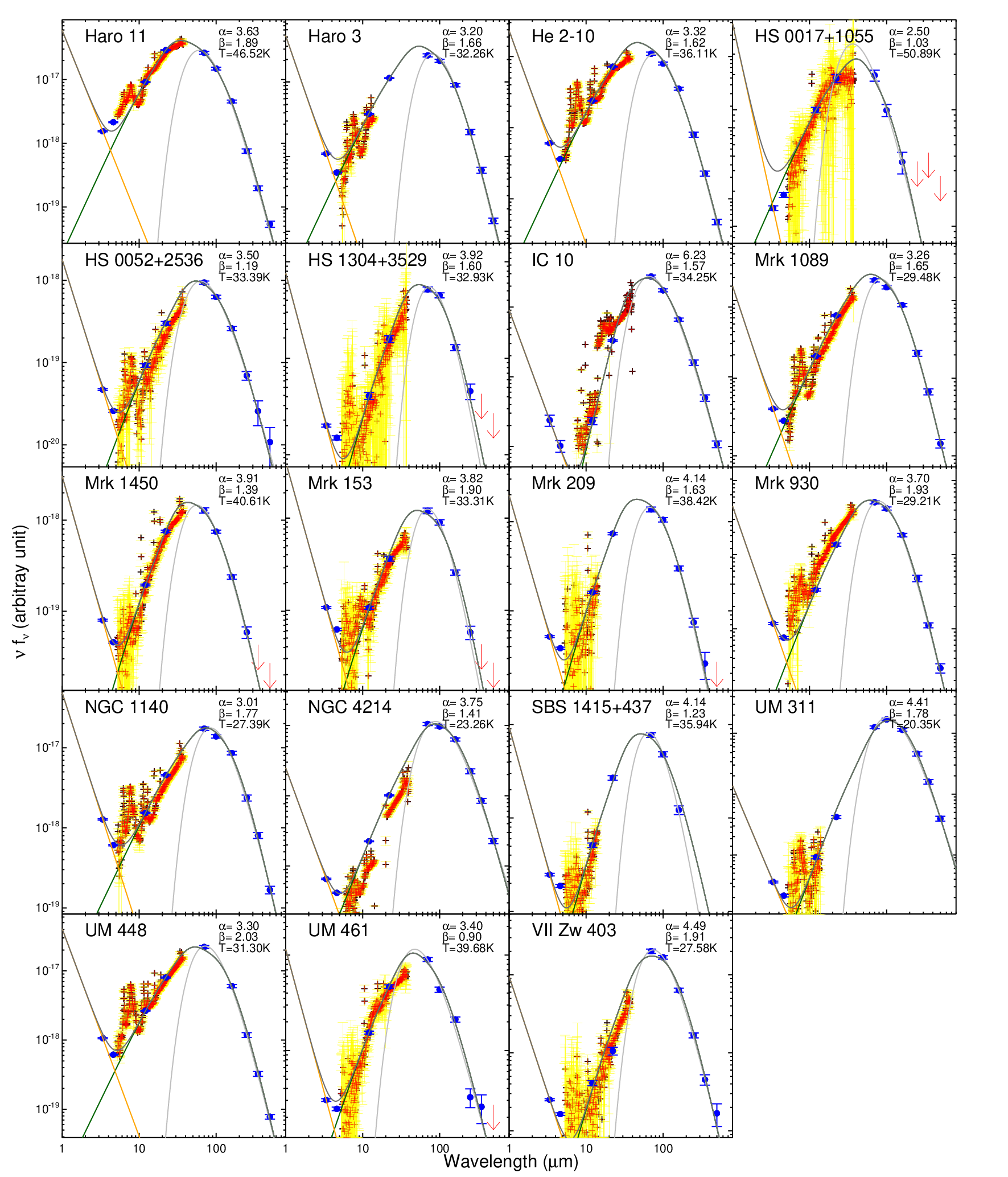} 
	\caption{
	    IR SEDs of the 19 DGS galaxies. Blue circles are for detections and
	    red down arrows are for upper limits. The available \spitzerirs
	    spectrum for each source (red crosses for the flux, yellow shadows
	    for the flux error) is also plotted. We first subtract the stellar
	    light in the SED by assuming a Rayleigh-Jeans tail scaled by the
	    WISE W1 band data point, then fit the data points at 8-1000$\mum$
	    with the modified blackbody + mid-IR power-law model (namely, CMC
	    fits, green lines). The final model SED is shown as dark grey
	    lines.  The $T$, $\alpha$, and $\beta$ parameters derived from the
	    CMC fits are indicated on the right up corner of each panel. In a
	    similar fashion as \cite{Remy-Ruyer2013}, a single modified
	    blackbody is also fit to the \herschel data points for comparison
	    (light grey line).
	} 
	\label{DGS_SED1} 
    \end{center} 
\end{figure*}

In Table~\ref{tab:list-template}, we list the basic information and fit
parameter values. In Figure~\ref{DGS_SED1}, we show the SED continuum fits for
the 19 DGS galaxies. We also present the results of single modified blackbody
fits (only on \herschel data) for comparison.

\begin{figure}[htp]
    \begin{center} 
	\includegraphics[width=0.9\hsize]{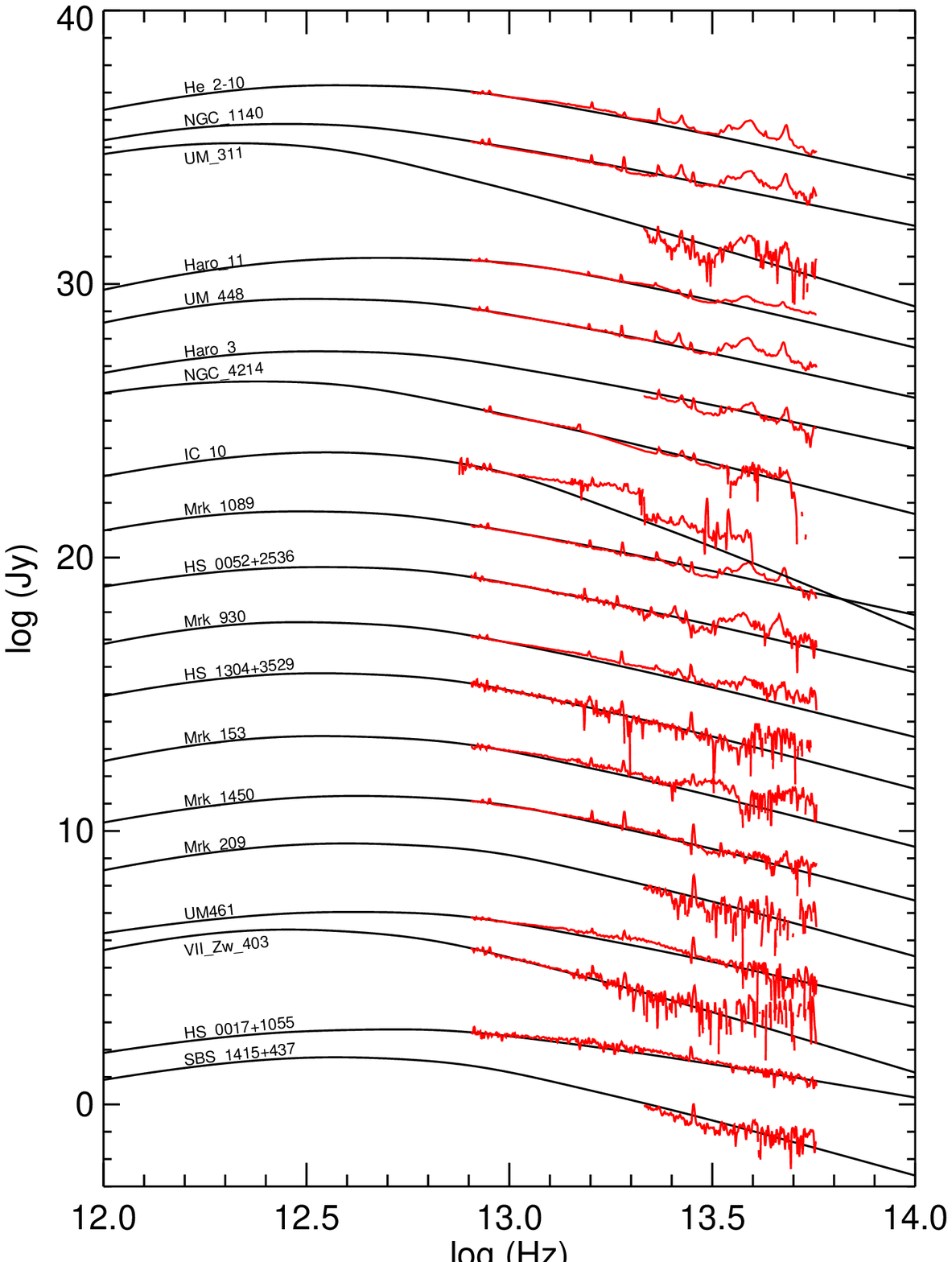} 
	\caption{
	    Full IR SEDs of the 19 DGS galaxies that have \spitzerirs
	    low-resolution spectra with high enough S/N, sequenced by their
	    metallicity with the lowest at the bottom. The \spitzer spectra are
	    scaled to match the continuum SEDs.
	} 
	\label{fig:dgs_template2} 
    \end{center} 
\end{figure}

Figure~\ref{fig:dgs_template2} shows the full infrared SEDs (after adding the
\spitzer spectra) for the 19 DGS galaxies.  The mid-IR continua
derived from the photometry fitting and those directly obtained from the mid-IR
spectra are consistent. The mid-infrared regions of these low-metallicity
objects present substantial forbidden line emissions and weak or no PAH
emission \citep[e.g.,][]{Wu2006, Wu2007}, which contribute little to the
continuum. For the final model SEDs of the 19 DGS galaxies, we discard the
mid-IR portion of the fit SED continua based on WISE and \herschel
photometry, and replace it with scaled \spitzer spectra.

\begin{deluxetable*}{lccccc}
    \tablecaption{Low-metallicity Galaxies Used to Derive the Templates\label{tab:list-template}}
\tablewidth{0pt}
\tablehead{
    \colhead{Source} & \colhead{12+log(O/H)} & \colhead{${\rm T_{dust}}$ (K)} 
    & \colhead{$\beta$} & \colhead{ $\alpha$ }
    & \colhead{Mid-IR Spectrum?} \\
    \colhead{(1)} & \colhead{(2)} & \colhead{(3)} & \colhead{(4)} & \colhead{(5)} & \colhead{(6)}
}
\startdata
Haro 11 	& 8.36$\pm$0.01  &     46.5$\pm$1.0  &     1.89  &      3.63    & good \\
Haro 3 		& 8.28$\pm$0.01  &     32.3$\pm$0.9  &     1.66  &      3.20    & limited \\
He 2-10 	& 8.43$\pm$0.01  &     36.1$\pm$0.7  &     1.62  &      3.32    & good \\
HS 0017+1055 	& 7.63$\pm$0.10  &     50.9$\pm$8.2  &     1.03  &      2.50    & good \\
HS 0052+2536 	& 8.04$\pm$0.10  &     33.4$\pm$2.1  &     1.19  &      3.50    & good \\
HS 1304+3529 	& 7.93$\pm$0.10  &     33.9$\pm$2.6  &     1.60  &      3.92    & good \\
IC 10 		& 8.17$\pm$0.03  &     34.3$\pm$0.8  &     1.57  &      6.23    & poor \\
Mrk 1089 	& 8.10$\pm$0.08  &     29.5$\pm$1.0  &     1.65  &      3.26    & good \\
Mrk 1450 	& 7.84$\pm$0.01  &     40.6$\pm$1.5  &     1.39  &      3.91    & good \\
Mrk 153 	& 7.86$\pm$0.04  &     33.3$\pm$2.3  &     1.90  &      3.82    & good \\
Mrk 209 	& 7.74$\pm$0.01  &     38.4$\pm$1.2  &     1.63  &      4.14    & limited \\
Mrk 930 	& 8.03$\pm$0.01  &     29.2$\pm$1.7  &     1.93  &      3.70    & good \\
NGC 1140 	& 8.38$\pm$0.01  &     27.4$\pm$1.0  &     1.77  &      3.01    & good \\
NGC 4214 	& 8.26$\pm$0.01  &     23.3$\pm$0.7  &     1.41  &      3.75    & limited \\
SBS 1415+437 	& 7.55$\pm$0.01  &     38.9$\pm$3.2  &     1.23  &      4.14    & limited \\
UM 311 		& 8.36$\pm$0.01  &     20.4$\pm$0.7  &     1.78  &      4.41    & limited \\
UM 448 		& 8.32$\pm$0.01  &     31.3$\pm$1.0  &     2.03  &      3.30    & good \\
UM 461 		& 7.73$\pm$0.01  &     40.0$\pm$2.0  &     0.90  &      3.40    & good \\
VII Zw 403 	& 7.66$\pm$0.01  &     27.6$\pm$1.8  &     1.91  &      4.49    & good
\enddata
\tablecomments{
    Col. (1): The galaxy name; Col. (2) The metallicity 12+log(O/H); Col. (3):
    The dust temperature derived from the CMC fit (see text); Col. (4): The
    far-IR emission index; Col. (5): The mid-IR power-law index; Col. (6): The
    existence of the mid-IR \spitzer spectrum: {\it good} -- the galaxy has
    full range $5-35\mum$ mid-IR spectrum; {\it limited} -- the galaxy has
    mid-IR spectrum with limited coverage; {\it poor} -- IC 10 is quite
    extended, as judged from its MIPS 24$\mum$ image, making the \spitzer
    spectrum a poor reflection of the overall mid-IR continuum.
}
\end{deluxetable*}

\subsection{Features of low-metallicity galaxy templates}\label{sec:dwarf-template-properties}

\begin{figure}[htp]
    \begin{center} 
	\includegraphics[width=0.9\hsize]{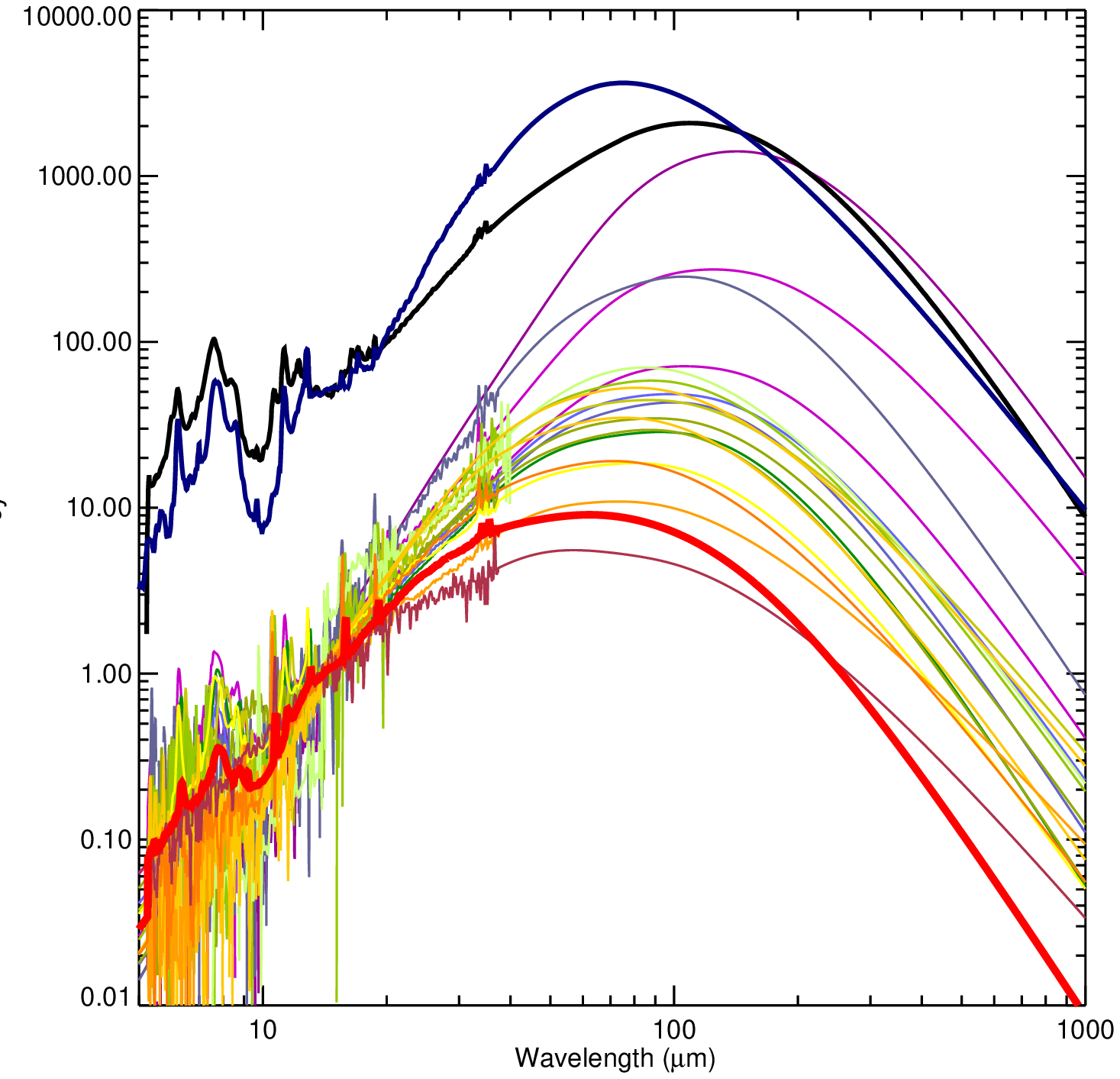} 
	\caption{
	    Family of full IR SEDs of the 19 DGS galaxies, normalized at
	    rest-frame 14$\mum$ (bottom group). We use thick red line to
	    highlight the SED of Haro 11.  For comparison, we present two
	    normal SF SED templates in \cite{Rieke2009}:
	    log$(L/L_{\odot})$= 12 (thick blue line), 11 (thick black line).
	} 
	\label{fig:dgs_template3} 
    \end{center} 
\end{figure}

Figure~\ref{fig:dgs_template3} compares the 19 low-metallicity templates with
the normal SF galaxy templates in \cite{Rieke2009}. The SEDs of
these dwarf galaxies show a lot of variation. Compared with the
solar-metallicity galaxies, the low-metallicity infrared SEDs derived in this
work tend to have the following features:
\begin{itemize}
    \item Higher far-IR dust temperature. For the low-metallicity galaxies, the
	typical dust temperature is $T_{\rm dust}=34\pm7.7$. Our value is
	similar to the result by \cite{Remy-Ruyer2013} ($T_{\rm dust}\sim
	32$K). Compared with the \herschel KINGFISH sample ($T_{\rm
	dust}\sim23$K, \citealt{Remy-Ruyer2013}), which contains more metal-rich
	environments, the dust in these low-metallicity systems is generally
	warmer;
    \item Steeply rising mid-IR continua. For low-metallicity galaxies in this
	work, the mid-IR continua slope $\alpha=3.8\pm0.8$, which is much
	larger than typical values in normal galaxies ($\alpha=2.0\pm0.5$ in
	\citealt{Casey2012}; 1.7-2.2 in \citealt{Blain2003});
    \item Weaker aromatic features. The contribution of aromatic features to
	the infrared SED of low-metallicity galaxies is substantially lower than
	normal galaxies. The weaker aromatic features with decreasing
	metallicity have been reported by many authors
	\citep[e.g.,][]{Engelbracht2005,Engelbracht2008,Madden2006}.
\end{itemize}

All of these features can be explained by a rich population of small (and/or hot) grains
in the low-metallicity environments (e.g., \citealt{Madden2006}). In addition,
some authors also report mm excess emission in these dwarf systems (e.g.,
\citealt{Galliano2003, Galliano2005,Galametz2009}). Since the origin of this
excess is not clear and its contribution to the infrared luminosity is tiny, we do
not consider it in this work.

\section{Star formation in dwarf galaxies}\label{sec:dwarf-sfr-calibration}

\subsection{Star Formation Determination}\label{sec:dwarf-sf-overall}

The star formation rate can be estimated by the 8-1000~$\mum$ infrared
emission, following the \cite{Kennicutt1998} star formation law,
\begin{equation}\label{eqn:sfr-law-k98}
    {\rm SFR}({\rm IR}, M_\odot/{\rm yr})  = 4.5\times10^{-44} L({\rm IR, erg/s}) ~.
\end{equation}
This relation is widely used for high-$z$ galaxies though it was originally
established for star-bursting galaxies. However, it is not clear if and how it
is valid for low-metallicity dwarf galaxies. 

In general, the star formation in a galaxy can be both dust-obscured and
dust-unobscured.  For unobscured star formation, tracers that probe direct
stellar light (e.g., the GALEX FUV at 0.153$\mum$) or ionized gas tracers
(e.g.,  H$\alpha$, Pa$\alpha$) are used.  For dwarf galaxies, \cite{Lee2009}
find FUV has a better performance than $H\alpha$ to trace the star formation,
since the latter tends to underpredict the total SFR relative to the FUV for low
luminosity systems. Therefore we use the FUV star formation law derived in
\cite{Salim2007}, a study that includes dwarf galaxies, to calculate the
dust-unobscured star formation rate:
\begin{equation}
    {\rm SFR}({\rm FUV}, M_\odot/{\rm yr}) = 1.08\times10^{-28} L_{\nu} ({\rm FUV, erg/s/hz}) ~.
\end{equation}
The dust-obscured star formation can be determined from the dust-processed
light at wavelengths where dust emission dominates (e.g., the 24 $\mum$
emission, the total infrared emission). We use MIPS 24$\mum$ star formation law
in \cite{Rieke2009}:
\begin{equation}
    {\rm SFR}(24\mum, M_\odot/{\rm yr}) = 2.02\times10^{-43} L(24\mum, {\rm erg/s}) ~.
\end{equation}
The final star formation rates of these dwarf galaxies are assumed to be a sum
of these two components, which will be compared with that derived from the
total infrared luminosity (Equation~\ref{eqn:sfr-law-k98}).

We collect GALEX FUV and MIPS 24$\mum$ data to derive the star formation.  All
of the 19 dwarf galaxies above have MIPS 24 $\mum$ observations, and 14/19 have
GALEX FUV observations. We retrieve their photometry from the catalog of
Spitzer Enhanced Imaging Products\footnote{
\url{http://irsa.ipac.caltech.edu/data/SPITZER/Enhanced/SEIP/overview.html}}
and the GALEX offical online catalog\footnote{
\url{http://galex.stsci.edu/GalexView/}}. IC10, NGC 4214, UM 311, VII Zw 403
are either too extended or have a close companion in their MIPS 24 $\mum$
images, thus they were removed from the comparison. GALEX FUV photometry can be
found for all sources with MIPS observations. 

\begin{figure}[htp]
    \begin{center} 
	\includegraphics[width=0.9\hsize]{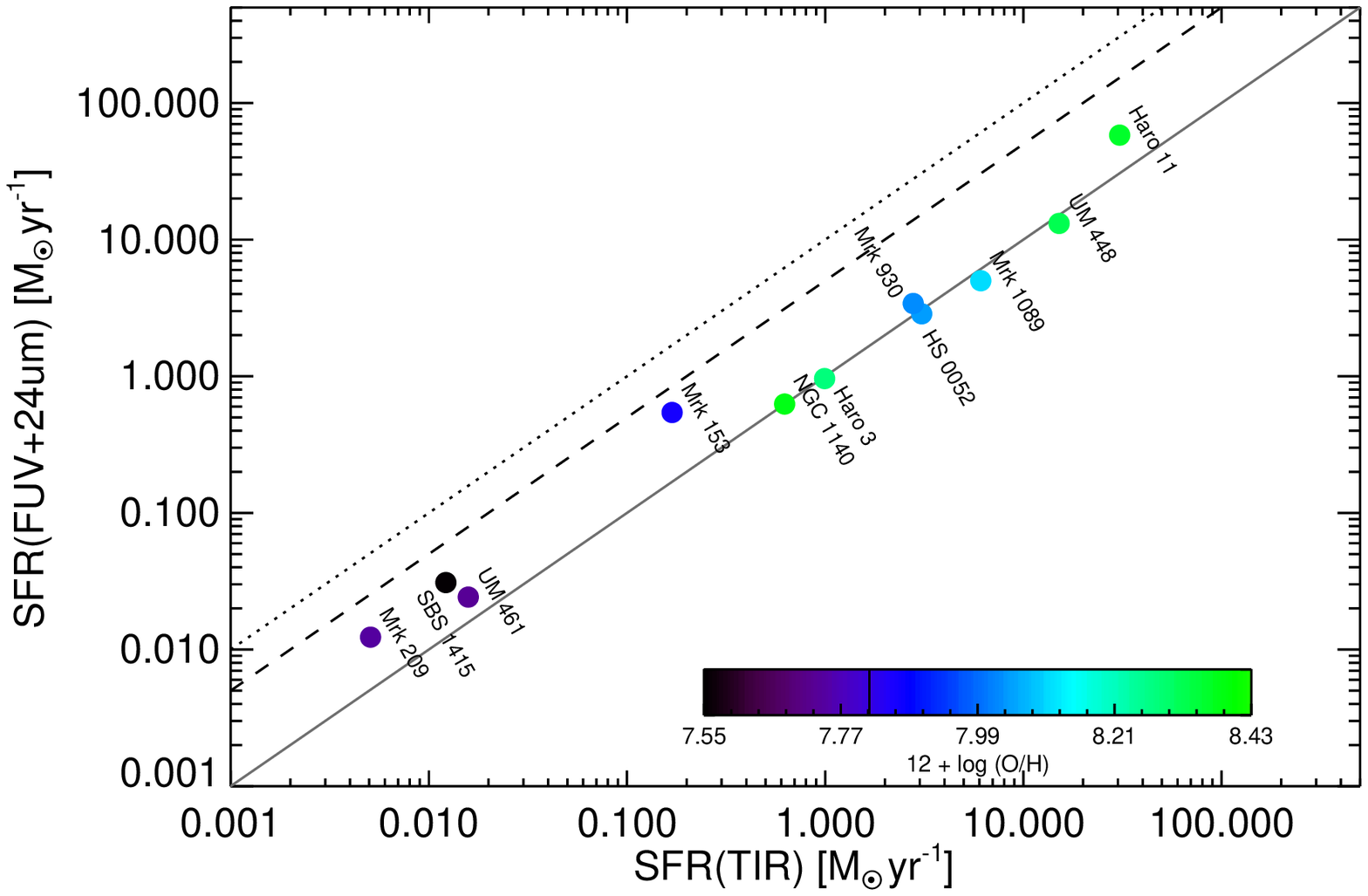} 
	\caption{
	    Comparison of star formation rates (SFRs) from the total infrared
	    luminosity (8-1000 $\mum$) and those from GALEX FUV + MIPS 24
	    $\mum$.  The colors code the metallicity 12+log(O/H). As a guide to
	    the eye, we also show the one-to-one (gray solid line), one-to-five
	    (gray dashed line), and one-to-ten (gray dotted line) lines.
	} 
	\label{fig:dgs_sfr} 
    \end{center} 
\end{figure}

In Figure~\ref{fig:dgs_sfr}, we compare the star formation rates derived from
the 8-1000$\mum$ infrared emission and those derived from the combined GALEX
FUV and MIPS 24$\mum$ emission. Within a three order-of-magnitude dynamical
    range, the star formation rates from these two approaches are generally
    consistent, without any obvious offset due to metallicity effect. Thus, we
    conclude the behavior of \cite{Kennicutt1998} $L_{\rm TIR}$ star formation
    law is similar to other star formation indicators for the population of
    dwarf galaxies, at least for those studied in this work, and not very
sensitive to metallicity.

\begin{deluxetable*}{l|lc|cccc}[htbl]
    \tablewidth{0pt}
    \tablecolumns{7}
    \tablecaption{The star formation rates of the dwarf galaxies\label{tab:dwarf-sfr}
		}
    \tablehead{ \colhead{Source} & 
		\colhead{12+(O/H)} &
		\colhead{$L_{\rm IR}$} &
		\colhead{SFR$_{\rm IR}$} & 
		\colhead{SFR$_{\rm FUV}$} & 
		\colhead{SFR$_{24\mum}$}  &
		\colhead{SFR$_{\rm FUV+24\mum}$} 
		\\
	        \colhead{ } & 
		\colhead{ } &
		\colhead{($10^{11} L_{\odot}$)} & 
		\colhead{($M_\odot/{\rm yr}$)} &
		\colhead{($M_\odot/{\rm yr}$)} & 
		\colhead{($M_\odot/{\rm yr}$)} & 
		\colhead{($M_\odot/{\rm yr}$)} 
		\\
	        \colhead{(1)} & \colhead{(2)} & \colhead{(3)} &
	        \colhead{(4)} & \colhead{(5)} & \colhead{(6)} & \colhead{(7)} 
	    }
\startdata
Haro 11		& 8.36$\pm$0.01	&  1.77     &  30.60    &   3.37  	&   56.06 	& 59.43 \\
Haro 3		& 8.28$\pm$0.01	&  0.057    &   0.99    &   0.22	&    0.82 	& 1.04	\\
He 2-10		& 8.43$\pm$0.01	&  0.061    &   1.05    &     --       &    1.13 	& --	\\
HS 0017+1055 	& 7.63$\pm$0.10	&  0.0095   &   0.16    &     --       &    0.26 	& --	\\
HS 0052+2536	& 8.04$\pm$0.10	&  0.18     &   3.07    &   1.57	&    1.88 	& 3.45	\\
HS 1304+3529	& 7.93$\pm$0.01	&  0.017    &   0.30    &     --       &    0.18 	& -- 	\\
IC 10		& 8.17$\pm$0.03	&  0.000055 &   0.00095 &     --       &       --	& -- 	\\
Mrk 1089	& 8.10$\pm$0.08	&  0.35     &   6.10    &   2.40	&    3.49 	& 5.89  \\
Mrk 1450	& 7.84$\pm$0.01	&  0.0030   &   0.053   &     --       &    0.064	& --	\\
Mrk 153		& 7.86$\pm$0.04	&  0.0098   &   0.17    &   0.67	&    0.12	& 0.79	\\
Mrk 209		& 7.74$\pm$0.01	&  0.00029  &   0.0051  &   0.012	&    0.0046 	& 0.017	\\
Mrk 930		& 8.03$\pm$0.01 &  0.16     &   2.78    &   1.13	&    2.70	& 3.83	\\
NGC 1140	& 8.38$\pm$0.01	&  0.036    &   0.62    &   0.49	&    0.32 	& 0.81	\\
NGC 4214	& 8.26$\pm$0.01	&  0.0040   &   0.070   &   0.076	&      --	& --	\\
SBS 1415+437 	& 7.25$\pm$0.01	&  0.00070  &   0.012   &   0.033	&    0.0094 	& 0.042	\\
UM 311		& 8.36$\pm$0.01	&  0.037    &   0.65    &   0.21	&      --	& --	\\
UM 448		& 8.32$\pm$0.01	&  0.88     &  15.15    &   2.19	&    11.76 	& 13.95 \\
UM 461		& 7.73$\pm$0.01	&  0.00091  &   0.016   &   0.0010	&    0.017  	& 0.028	\\
VII Zw 403	& 7.66$\pm$0.01	&  0.00016  &   0.0028  &   0.0065	&      --        & --	
\enddata
\tablecomments{ 
    Objects with `--' in Column (5), (6) do not have corresponding observations
    or are not point-sources to be included in the catalogs (see text).
}
\end{deluxetable*}

\subsection{Star Formation Law of Haro 11}\label{sec:dwarf-sf-haro11}

As discussed in Section~\ref{sec:templates}, Haro 11 was selected as the
candidate galaxy template for high-$z$ galaxies. With ample observations
presented in the literature, its star formation rate can be estimated using
various star-formation indicators, including the 8-1000 $\mum$ infrared
luminosity, MIPS 24 $\mum$ emission, H$\alpha$ emission-line luminosity, and
FUSE/GALEX FUV luminosity. Following the calibrations listed above, we estimate
its star formation rates and summarize these results in
Table~\ref{tab:haro11-sfr}. 

\begin{deluxetable*}{lcccc}
    \tablecaption{Star Formation Rate of Haro 11\label{tab:haro11-sfr}}
\tablewidth{0pt}
\tablehead{
    \colhead{Method} & \colhead{Luminosity ($L_\odot$)$^a$} & \colhead{ SFR ($M_\odot/{\rm yr}$)} 
    & \colhead{SFR Law} & \colhead{Reference}
}
\startdata
$L_{\rm IR}$		& $1.8\times10^{11}$	&     30.6  & \cite{Kennicutt1998}  &      this work   \\
MIPS 24 $\mum$  	& $6.7\times10^{10}$	&     56.1  & \cite{Rieke2009}      &      this work   \\
H$\alpha$ (Fabry-Perot)$^b$ 	& $1.3\times10^{9}$	&     39.1   & \cite{Kennicutt1994}  &      \cite{Ostlin1999} \\
H$\alpha$ (HST image)$^b$ 	& $8.7\times10^{8}$	&     25.7   & \cite{Kennicutt1994}  &      \cite{Ostlin2009} \\
GALEX FUV (0.153$\mum$)$^c$ 	& $1.6\times10^{10}$ 	&     3.4   & \cite{Salim2007}      &      this work   \\
FUSE FUV (0.115$\mum$)$^c$ 	& $2.2\times10^{10}$ 	&     4.5   & \cite{Kennicutt1998}  &      \cite{Grimes2007}  
\enddata
\tablecomments{
    $^a$ All luminosities are scaled to the distance~92.1 Mpc
    \citep{Bergvall2006}.  
    $^b$ We corrected the $H\alpha$ extinction based on the observed Balmer
    decrement $H\alpha/H\beta=4.08$ \citep{Bergvall2002} with the assumption of
    an intrinsic ratio 2.85 and a \cite{Calzetti2001} extinction law. The
    attenuation at V-band is estimated to be $A_V=0.95$, suggesting the
    intrinsic H$\alpha$ flux would be a factor of 1.34 larger than the
    observed.
    $^c$ For the UV band, since we do not make extinction corrections, the
    SFR(FUV) listed here are only for unobscured star formation.
}
\end{deluxetable*}

\begin{figure}[htbp]
    \begin{center} 
	\includegraphics[width=0.9\hsize]{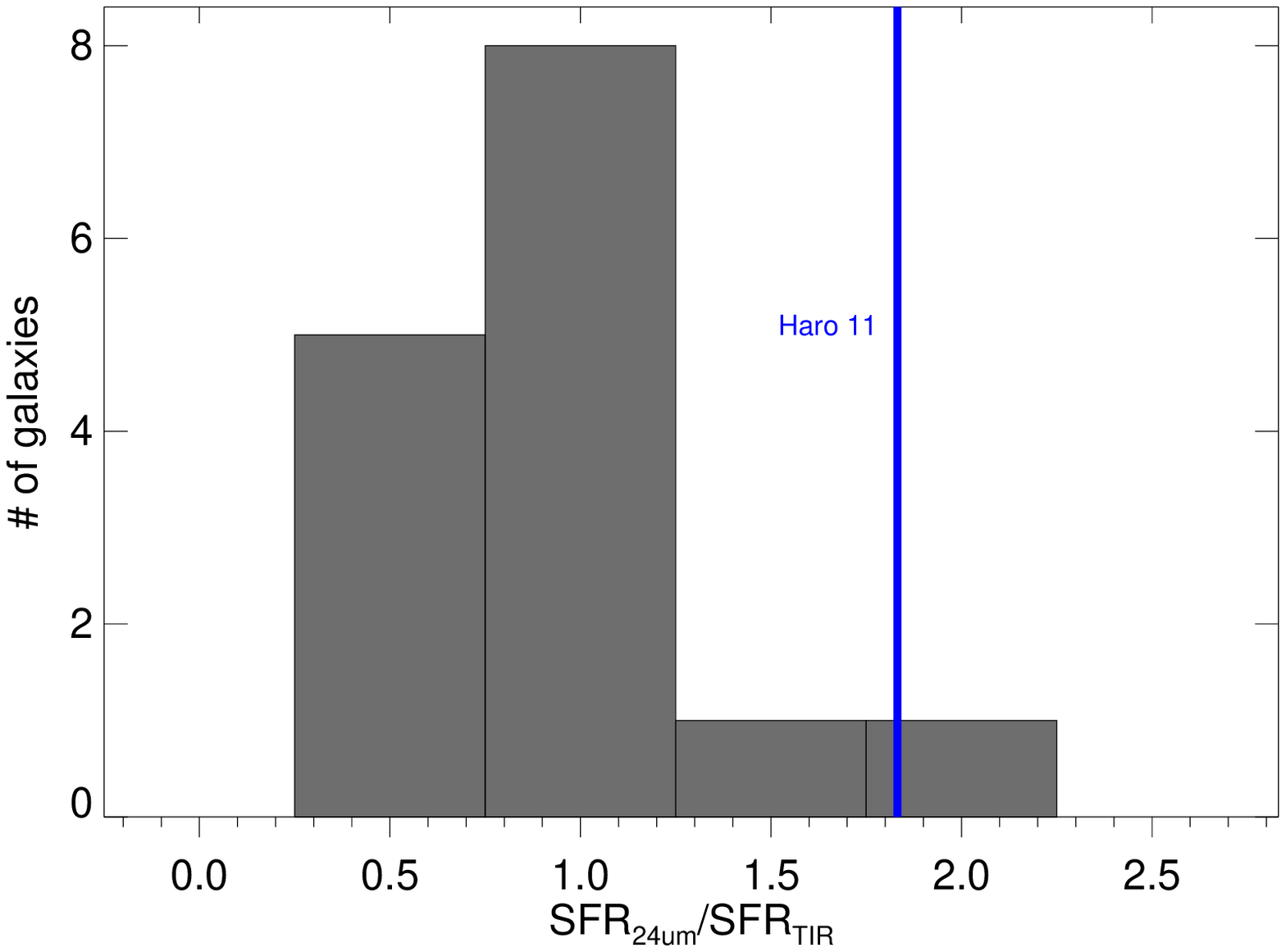} 
	\caption{ 
	    Histogram of the ratio between the 24$\mum$-based SFR and
	    $L_{\rm IR}$-based SFR of dwarf galaxies. Haro 11 has a
	    SFR(24$\mum$)/SFR(TIR)$~$1.82 as denoted in the blue line.
	} 
	\label{fig:haro11_hist} 
    \end{center} 
\end{figure}

The SFR determined from the 24~$\mum$ emission is $\sim1.8$ times higher than that
from either $L_{\rm IR}$ or $L(H\alpha)$. This discrepancy is a direct
consequence of its hot SED \citep{Engelbracht2008} and the resulting large
fraction of its $L_{\rm IR}$ emitted at 24~$\mum$ (see
Figure~\ref{fig:haro11_hist})\footnote{Another galaxy with
SFR(24$\mum$)/SFR(FIR)$>1.25$ in our low-metallicty galaxy sample is Mrk 153,
which hosts an AGN that boosts the mid-IR emission.}.  Given the close
agreement in the SFRs from H$\alpha$ and $L_{\rm IR}$, we conclude that Haro 11
falls above the trend line in Figure~\ref{fig:dgs_sfr} because FUV+24~$\mum$
overestimates the SFR and $L_{\rm IR}$ gives a valid estimate.  Moreover, since
Haro 11 presents very young stellar populations \citep{Adamo2010}, the
contamination of emission from old stars to the far infrared emission is
negligible, which makes the $L_{\rm IR}$, following the \cite{Kennicutt1998} SF
law, a robust tracer of the obscured star formation

We use the FUV luminosity based on GALEX observation and assume the star
formation law in \cite{Salim2007} to estimate the unobscured star formation.
The final adopted star formation rate of Haro 11, which is 34.0 $M_\odot/{\rm
yr}$, is based on a combination of GALEX FUV emission and the 8-1000 $\mum$
infrared emission.  We can finally relate the infrared luminosity and FUV
luminosity of Haro 11 to its total star formation rate, which includes both the
obscured and the unobscured, as

\begin{align}
    {\rm SFR}_{\rm Haro~11} (M_\odot/{\rm yr}) & = 5.00\times10^{-44}  L{(\rm IR, erg/s)} \\
                                                      & = 1.92\times10^{-10} L{(\rm IR, L_\odot)} 
\end{align}
This is the star formation law used for the Haro 11 template in the main part
of this paper.

\end{document}